\documentclass[titlepage,pallis,twoside]{wpallis}
\usepackage{fancyh}
\usepackage[l]{floatflt}
\usepackage{epsfig}
\usepackage{amssymb}
\usepackage{latexsym}
\usepackage{times}
\usepackage{slashed,cite}
\usepackage{upgreek}
\numberwithin{equation}{section}
\usepackage{rotating}
\usepackage{amsmath}
\usepackage{amsfonts}
\usepackage{amsbsy}
\usepackage{amscd}
\usepackage{bbm}
\usepackage{float}

\newcommand{\bal}{\begin{align}}
\newcommand{\eal}{\end{align}}
\newcommand{\beqs}{\begin{subequations}}
\newcommand{\eeqs}{\end{subequations}}
\newcommand{\eec}{\end{center}}
\newcommand{\bec}{\begin{center}}

\newcommand{\eem}{\end{matrix}}
\newcommand{\bem}{\begin{matrix}}
\newcommand{\eeq}{\end{equation}}
\newcommand{\beq}{\begin{equation}}
\newcommand{\ba}{\begin{array}}
\newcommand{\ea}{\end{array}}
\newcommand{\bea}{\begin{eqnarray}}
\newcommand{\eea}{\end{eqnarray}}
\newcommand{\baq}{\begin{eqnarray}}
\newcommand{\eaq}{\end{eqnarray}}

\newcommand\eqs[2]{Eqs.~(\ref{#1}) and (\ref{#2})}

\newcommand\eqss[3]{Eqs.~(\ref{#1}), (\ref{#2}) and (\ref{#3})}

\newcommand{\ftn}{\footnotesize}

\newcommand{\ssz}{\scriptsize}

\newcommand{\GeV}{{\mbox{\rm GeV}}}

\newcommand{\erfi}{{\mbox{\rm erfi}}}

\newcommand{\sFref}[2]{Fig.~\ref{#1}-{\ftn\sf ({#2})}}
\newcommand{\sEref}[2]{Eq.~(\ref{#1}{\ftn\sf {#2}})}
\newcommand{\crefs}[1]{Refs.~\cite{#1}}
\newcommand{\etal}{{\it et al.\/}}

\def\to{\rightarrow}

\def\lf{\left(}
\def\rg{\right)}
\newcommand\vev[1]{\langle {#1} \rangle}

\newcommand{\Vhi}{\ensuremath{\widehat V_{\rm CI}}}
\newcommand{\Ei}{\ensuremath{{\rm Ei}}}

\newcommand{\Vjhi}{\ensuremath{V_{\rm CI}}}
\newcommand{\Hhi}{\ensuremath{\widehat H_{\rm CI}}}

\newcommand{\Khi}{\ensuremath{K}}

\newcommand{\Ns}{\ensuremath{{\what N_\star}}}
\newcommand{\ck}{\ensuremath{c_{\rm K}}}
\newcommand{\cks}{\ensuremath{c_{1\rm K}}}

\newcommand{\mP}{\ensuremath{m_{\rm P}}}

\newcommand{\Qef}{\ensuremath{\Lambda_{\rm UV}}}

\def\openone{\leavevmode\hbox{\small1\kern-3.8pt\normalsize1}}
\newcommand{\dV}{\ensuremath{\Delta\widehat V_{\rm CI}}}

\newcommand{\frs}{\ensuremath{f_{1\cal R}}}

\newcommand{\fr}{\ensuremath{f_{\cal R}}}

\newcommand{\fk}{\ensuremath{f_{\rm K}}}
\newcommand{\hk}{\ensuremath{F_{\rm K}}}
\newcommand{\hr}{\ensuremath{F_{\cal R}}}
\newcommand{\he}{\ensuremath{F_{p}}}

\newcommand{\kx}{\ensuremath{k_S}}

\newcommand{\ca}{\ensuremath{c_{\cal R}}}
\newcommand{\fa}{\ensuremath{F_{1S}}}
\newcommand{\fb}{\ensuremath{F_{2S}}}

\newcommand{\ks}{\ensuremath{k_\star}}
\newcommand{\ns}{\ensuremath{n_{\rm s}}}
\newcommand{\as}{\ensuremath{a_{\rm s}}}
\newcommand{\As}{\ensuremath{A_{\rm s}}}
\newcommand{\rw}{\ensuremath{r_{0.002}}}
\newcommand{\rs}{\ensuremath{r_{\mathcal{R}\rm K}}}
\newcommand{\rcc}{\ensuremath{\mathcal{R}}}
\newcommand{\rce}{\ensuremath{\widehat{\mathcal{R}}}}
\newcommand{\Ve}{\ensuremath{\widehat{V}}}

\newcommand{\dphi}{\ensuremath{\what{\delta\phi}}}
\newcommand{\dph}{\ensuremath{\delta\phi}}

\newcommand{\what}{\ensuremath{\widehat}}

\def\bbet{{\bar\beta}}
\def\al{{\alpha}}

\def\th{{\theta}}

\newcommand{\Trh}{\ensuremath{T_{\rm rh}}}
\newcommand{\sg}{\ensuremath{\phi}}

\newcommand{\sgx}{\ensuremath{\phi_\star}}
\newcommand{\sgf}{\ensuremath{\phi_{\rm f}}}
\newcommand{\sgmax}{\ensuremath{\phi_{\rm max}}}

\newcommand{\ld}{\ensuremath{\lambda}}

\newcommand{\Ld}{\ensuremath{\Lambda}}

\newcommand{\Dex}{\ensuremath{\Delta_{\rm max\star}}}

\newcommand{\se}{\ensuremath{\widehat \phi}}
\newcommand{\sex}{\ensuremath{\widehat{\phi}_\star}}
\newcommand{\sef}{\ensuremath{\widehat{\phi}_{\rm f}}}
\newcommand{\geu}{\ensuremath{\widehat g}}
\newcommand{\eph}{\ensuremath{\widehat \epsilon}}
\newcommand{\ith}{\ensuremath{\widehat \eta}}

\newcommand{\nb}{\ensuremath{N_{S}}}

\def\Ka{K\"{a}hler potential}
\def\Km{K\"{a}hler manifold}
\def\Kaa{K\"{a}hler~}

\def\FHI{nMI~}
\def\bcp{{\sc\small Bicep2}/{\it Keck Array}}
\newcommand{\plk}{{\it Planck}}

\renewcommand{\figurename}{\bf\scshape Fig.}
\renewcommand{\tablename}{\bf\scshape Table}
\renewcommand{\refname}{{\bf\scshape References}}

\begin{document}


\title{\bf\scshape Kinetically Modified Non-Minimal Inflation With Exponential Frame Function}

\author{\large\bfseries\scshape  C. Pallis}
\address[] {\sl Department of Physics, University of Cyprus, \\ P.O. Box 20537,
CY-1678 Nicosia, CYPRUS \\  {\sl e-mail address: }{\ftn\tt
cpallis@ucy.ac.cy}}



\begin{abstract}

\noindent {\small \bf\scshape Abstract} \\ \par We consider
\emph{Supersymmetric} (SUSY) and non-SUSY models of chaotic
inflation based on the $\phi^n$ potential with $n=2$ or $4$. We
show that the coexistence of an exponential non-minimal coupling
to gravity $\fr=e^{\ca\phi^{p}}$ with a kinetic mixing of the form
$\fk=\ck\fr^m$ can accommodate inflationary observables favored by
the \plk\ and \bcp\ results for $p=1$ and $2$, $1\leq m\leq15$ and
$2.6\cdot10^{-3}\leq\rs=\ca/\ck^{p/2}\leq1,$ where the upper limit
is not imposed for $p=1$. Inflation is of hilltop type and it can
be attained for subplanckian inflaton values with the
corresponding effective theories retaining the perturbative
unitarity up to the Planck scale. The supergravity embedding of
these models is achieved employing two chiral gauge singlet
supefields, a monomial superpotential and several
(semi)logarithmic or semipolynomial \Ka s.
\\ \\ {\scriptsize {\sf PACS codes: 98.80.Cq, 11.30.Qc, 12.60.Jv,
04.65.+e}

}

\end{abstract}\pagestyle{fancyplain}

\maketitle

\rhead[\fancyplain{}{ \bf \thepage}]{\fancyplain{}{\sl Kinetically
Modified nMI With Exponential Frame Function}}
\lhead[\fancyplain{}{\sl C. Pallis}]{\fancyplain{}{\bf \thepage}}
\cfoot{}

\tableofcontents\vskip-1.3cm\noindent\rule\textwidth{.4pt}\\

\section{Introduction} \label{intro}

In a recent paper \cite{nMkin} we show that the consideration of a
monomial potential of the type
\beq \label{Vn}\Vjhi(\sg)=\ld^2\sg^n/2^{n/2},\eeq
for the inflaton $\sg$ in conjunction with a non-minimal coupling
function
\beq \label{fr} \fr(\sg)=1+\ca\sg^{n/2},  \eeq
between $\sg$ and the Ricci scalar $\rcc$, and a non-minimal
kinetic mixing of the form
\beq\label{fk} \fk(\sg)=\ck\fr^m,\eeq
gives rise to a novel type of \emph{non-minimal (chaotic)
inflation} ({\sf\ftn nMI}) \cite{old, sm1, nmi, atroest, nmH,
linde1, quad} named \emph{kinetically modified}. The inflationary
observables of these models can become impressively compatible
with the latest data \cite{plin,gwsnew} provided that the
parameter $\rs=\ca/\ck^{n/4}$ is adjusted to natural values. Most
notably, the resulting tensor-to-scalar ratio not only respects
the upper bound imposed by \plk\ \cite{plin} but also can be
confined in the $1-\sigma$ margin of the \bcp\ data \cite{gwsnew}.
Moreover, the corresponding effective theories are stable against
corrections from higher order terms and unitarity safe up to the
Planck scale for $\rs\leq1$ and any $n$ -- cf.
\cref{cutoff,riotto}. In short, the simple, predictive and
well-motivated models based on $\Vjhi$ of \Eref{Vn}, which are by
now observationally \cite{plin} excluded in their minimal form for
$n\geq2$, can be totally revitalized and probed in the near future
\cite{ictp} within our scheme at the cost of two more parameters
($m$ and $\rs$) for fixed $n$.


Trying to test the applicability of our proposal, we here continue
developing SUSY and non-SUSY inflationary models relied on the
same structure. Namely, we keep $\Vjhi$ in \Eref{Vn} and the
functional dependence of $\fk$ on $\fr$ in \Eref{fk} whereas we
replace the polynomial $\fr$ in \Eref{fr} with an exponential one
introducing one more parameter $p$, i.e., we here adopt
\beq \label{fre} \fr(\sg)=e^{\ca\sg^{p}}
\>\>\mbox{with}\>\>\ca>0\,. \eeq
More specifically, we present inflationary solutions employing
$n=2$ and $4$ in \Eref{Vn}, $p=1$ and $2$ in \Eref{fre} and
various $m$'s in \Eref{fk}. The emerging models are drastically
different from the original ones in \cref{nMkin} -- see also
\crefs{jhep, nMHkin, var} -- since they support exclusively
inflation of hilltop type \cite{lofti}. However, they share
similar observational and theoretical advantages with the initial
ones.

Our scheme can be also incarnated in the context of
\emph{Supergravity} ({\sf\ftn SUGRA}) by adopting two chiral
superfields, a monomial superpotential, $W$, and various \Ka s,
$K$, which may cooperate with it. In particular, we specify ten
different $K$'s using two possible functional forms for the
inflaton field and two possible stabilization methods
\cite{linde1, su11} for the accompanying field. A common feature
of all these $K$'s is the presence of a shift symmetric term for
taming \cite{nan, shift, lindeExp} the well-known $\eta$ problem
of inflation within SUGRA -- such continuous symmetries appear
naturally at tree-level due to the underlying discrete modular
symmetries of the full string theory as pointed out in
\cref{string}. The non-minimal gravitational coupling and kinetic
mixing of the inflaton arise as violations of this shift symmetry
-- for similar attempts see \cref{lee}.

The exponential frame function and kinetic mixing  are used for
the dilaton in the low-energy effective string theories -- see
\cref{dilaton} where mostly negative $\ca$'s are considered. The
resulting potential has the form $\sg^ne^{-2\ca\sg^p}$ and
resembles the one met in the models of logamediate inflation
\cite{barrow, review1}. However, the non-minimal kinetic mixing
and gravitational coupling in \eqs{fk}{fre} make the canonically
normalized inflaton different than $\sg$ and so the inflationary
dynamics in our case is clearly distinguishable. In the SUGRA
framework only some subclasses of the set of models introduced
here are investigated in \crefs{nan,lindeExp}. Most notably, in
\cref{nan} the cases with $n=1,3/2, 2$, $m=1$ and $p=1,2$ are
investigated adopting one of the $K$'s suggested also here. The
consideration of $n=4$ and $m>1$ together with the analysis of the
small-field behavior of the models consist our main improvements
in the present work.

Below we first, in \Sref{nonsusy}, establish our setting in a
non-SUSY framework and then, in \Sref{fhim}, we outline its
possible implementations in SUGRA. The resulting inflationary
models are tested against observations in \Sref{fhi}. Our
conclusions are finally summarized in Sec.~\ref{con}. Throughout
the text, we use units where the reduced Planck scale $\mP =
2.433\cdot 10^{18}~\GeV$ is set to be unity, the subscript of type
$,z$ denotes derivation \emph{with respect to} ({\small\sf w.r.t})
the field $z$ and charge conjugation is denoted by a star ($^*$).

\section{non-SUSY Framework}\label{nonsusy}

According to its definition \cite{nMkin}, kinetically modified nMI
is formulated in the \emph{Jordan frame} ({\sf\ftn JF}) where the
action of $\phi$ is given by
\beq \label{action1} {\sf  S} = \int d^4 x \sqrt{-\mathfrak{g}}
\left(-\frac12\fr\rcc +\frac12\fk g^{\mu\nu}
\partial_\mu \sg\partial_\nu \sg-
\Vjhi(\sg)\right), \eeq
where the involved functions $\Vjhi,\fk$ and $\fr$ are defined in
\eqss{Vn}{fk}{fre} respectively. Furthermore, $\mathfrak{g}$ is
the determinant of the background Friedmann-Robertson-Walker
metric $g^{\mu\nu}$ with signature $(+,-,-,-)$. The \emph{vacuum
expectation value} ({\ftn\sf v.e.v}) of $\sg$ is $\vev{\sg}=0$,
and the validity of ordinary Einstein gravity at low energies is
guaranteed since $\vev{\fr} =1$.

By performing a conformal transformation \cite{old, nmi} according
to which we define the \emph{Einstein frame} ({\sf\ftn EF}) metric
$\geu_{\mu\nu}=\fr\,g_{\mu\nu}$ we can write ${\sf S}$ in the EF
as follows
\beq {\sf  S}= \int d^4 x
\sqrt{-\what{\mathfrak{g}}}\left(-\frac12
\rce+\frac12\geu^{\mu\nu} \partial_\mu \se\partial_\nu \se
-\Vhi(\se)\right), \label{action} \eeq
where hat is used to denote quantities defined in the EF. The EF
canonically normalized field $\se$ and potential $\Vhi$ are given
as functions of the initial field, $\sg$, through the relations
\beq\label{VJe}
\frac{d\se}{d\sg}=J=\sqrt{\frac{\fk}{\fr}+{3\over2}\left({f_{\cal
R,\sg}\over
\fr}\right)^2}=\sqrt{{\ck\over\fr^{1-m}}+\frac32p^2\ca^2\sg^{2(p-1)}}\hspace*{0.3cm}\mbox{and}\hspace*{0.3cm}\Vhi=
\frac{\Vjhi}{\fr^2}={\ld^2\sg^n\over2^{\frac{n}{2}}
e^{2\ca\sg^p}}\hspace*{0.3cm}\eeq
where \eqs{Vn}{fre} are considered. The resulting $\Vhi$
represents an almost Gaussian profile clearly disguisable from its
shape in the models of the pure \cite{old, nmi, atroest} or
kinetically modified \cite{nMkin, jhep, nMHkin, var} nMI  where an
almost flat plateau emerges in EF -- see \Sref{num} too. The
parametrization of $\fk$ in \Eref{fk} assists us to simplify the
derived $J$ in \Eref{VJe}. However, the variation of $m$ allows us
to scan a very wide range of the parametric space.

To determine transparently it, we perform a rescaling
$\sg=\tilde\sg/\sqrt{\ck}$. Then, \Eref{action1} preserves its
form replacing $\sg$ with $\tilde\sg$ and $\fk$ with $\fr^m$ where
$\fr$ and $\Vhi$ take, respectively, the forms
\beq\label{frVrs} \fr=e^{\rs\tilde\sg^{p}}\>\>\>\mbox{and}\>\>\>
\Vhi=2^{-n/2}\ld^2\fr^{-2}\tilde\sg^{n}\ck^{n/2}\>\>\>\mbox{with}\>\>\>\rs=\ca
\ck^{-p/2}\,.\eeq
Therefore, we expect that our scenario depends only on $\rs$ and
$\ld^2/\ck^{n/2}$ for fixed $n, p$ and $m$.


\section{Supergravity Framework} \label{fhim}

The inflationary model introduced above (in the non-SUSY
framework) can be also implemented in the context of SUGRA as
established in \Sref{fhim1} and verified in \Sref{fhim2}.

\subsection{Possible Embeddings}  \label{fhim1}

In \Sref{fhi1a} we outline the basics of the SUGRA regime. Then,
we specify a variety of logarithmic or semi-logarithmic -- in
\Sref{fhi2} -- and semi-polynomial -- in \Sref{fhi3} -- $K$'s
obeying a number of enhanced symmetries pointed out in
\Sref{fhi4}.

\subsubsection{The General Set-up}  \label{fhi1a}

The SUGRA versions of our model can be easily constructed if we
use two gauge singlet chiral superfields, i.e., $z^\al=\Phi, S$,
with $\Phi$ ($\al=1$) and $S$ ($\al=2)$ being the inflaton and a
``stabilizer'' field respectively. The EF action for $z^\al$'s
within SUGRA \cite{linde1} can be written as
\beqs \beq\label{Saction1} {\sf S}=\int d^4x \sqrt{-\what{
\mathfrak{g}}}\lf-\frac{1}{2}\rce +K_{\al\bbet}
\geu^{\mu\nu}\partial_\mu z^\al \partial_\nu z^{*\bbet}-\Ve\rg
\eeq
where summation is taken over the scalar fields $z^\al$, denoted
by the same superfield symbol, $K_{\al\bbet}=K_{,z^\al
z^{*\bbet}}$ and
$K^{\al\bbet}K_{\bbet\gamma}=\delta^\al_{\gamma}$. Also $\Ve$ is
the EF F--term SUGRA potential given by
\beq \Ve=e^{\Khi}\left(K^{\al\bbet}(D_\al W) (D^*_\bbet
W^*)-3{\vert W\vert^2}\right)\>\>\mbox{with}\>\>\>D_\al
W=W_{,z^\al} +K_{,z^\al}W\,.\label{Vsugra} \eeq \eeqs

Defining the inflationary trajectory by the constraints
\beq \label{inftr} S=\Phi-\Phi^*=0,\>\>\mbox{or}\>\>\>s=\bar
s=\th=0\eeq
if we express $\Phi$ and $S$ according to the parametrization
\beq \Phi=\:{\phi\,e^{i
\th}}/{\sqrt{2}}\>\>\>\mbox{and}\>\>\>S=\:(s +i\bar
s)/\sqrt{2}\,,\label{cannor} \eeq
we can derive $V_{\rm CI}$ in \Eref{Vn}, in the flat limit, by the
superpotential
\beq \label{Wn} W=\ld S\Phi^{n/2},\eeq
since $\Vjhi=|W_{,S}|^2$. The form of $W$ can be uniquely
determined if we impose an $R$ symmetry, under which $S$ and
$\Phi$ have charges $1$ and $0$, and a global $U(1)$ symmetry with
assigned charges $-1$ and $2/n$ for $S$ and $\Phi$. The latter is
violated though in the proposed $K$'s below. The same $W$ is
considered also in \cref{nan}. Here we focus on $n=2$ and $4$
which are mainly encountered in concrete particle models. E.g.,
for $n=2$ this  $W$ is employed in the first paper in \cref{quad}
and for $n=4$ this  $W$ is adopted in \cref{tony} and the last
paper in \cref{linde1}.

On the other hand, $\Vhi$ in \Eref{VJe} can be derived from
\Eref{Vsugra} by conveniently choosing $K$. The consideration of
$S$ facilitates this aim, since only one term from $\Ve$ survives
along the path in \Eref{inftr}, which reads
\beq \label{1Vhio}\Vhi=\Ve(\th=s=\bar s=0)=e^{K}K^{SS^*}\,
|W_{,S}|^2\,.\eeq
The selected $K$'s should incorporate $\fr$ in \Eref{fre} and
$\fk$ in \Eref{fk}. To this end we introduce the functions
\beqs\beq \label{hr}
\hr(\Phi)=\exp\he(\Phi)\>\>\>\mbox{with}\>\>\>\he(\Phi)=
2^{p/2}\ca\Phi^{p}\>\>\>\mbox{and}\>\>\>\hk=(\Phi-\Phi^*)^2\,.
\eeq
Here \hr\ is an holomorphic function reducing to $\fr$, along the
path in \Eref{inftr}, and $\hk$ is a real function which assists
us to incorporate the non-canonical kinetic mixing generating by
$\fk$ in \Eref{fk}. Indeed, $\hk$ lets intact $\Vhi$, since it
vanishes along the trajectory  in \Eref{inftr}, but it contributes
to the normalization of $\Phi$. For the same reason, terms of the
form $k_{\rm KK}\hk^2+k_{S\rm K}|S|^2\hk$ are practically
irrelevant in our analysis and are not included for simplicity in
the expression of the $K$'s given below. We also include in $K$'s
the typical kinetic term for $S$, in terms of the functions
\beq \fa=|S|^2-\kx{|S|^4}\>\>\>\mbox{or}\>\>\>\fb=1+|S|^2/\nb\,,
\label{Fs}\eeq\eeqs
where we consider the next-to-minimal term in $\fa$ for stability
reasons \cite{linde1}. Alternatively, we can assume that $S$ is a
nilpotent superfield \cite{nil} and the second term in the
definition of $\fa$ can be avoided.

\subsubsection{(Semi)Logarithmic \Kaa\ Potentials}\label{fhi2}

The conventional embedding of a non-minimal coupling within SUGRA
is realized \cite{linde1} using a logarithmic $K$ which include it
in its argument. Applying this recipe in our case we arrive at
\beqs\beq
K_1=-3\ln\big((\hr+\hr^*)/2-\fa/3+\ck(\hr+\hr^*)^m\hk/6\cdot2^m
\big)\,,~~~~~\label{K1}\eeq
where we can easily recognize the similarities with the $K$'s
introduced in \cref{nMkin}. Using the reasoning of \cref{var} and
insisting on integer prefactors for the logarithms, to avoid any
relevant tuning, we can enumerate other six semi-logarithmic $K$'s
which yield identical results, i.e.,
\bea K_2&=&-3\ln\big((\hr +\hr^*)/2 -
\fa/3\big)-\ck(\hr+\hr^*)^{m-1}\hk/2^m\,,\label{K2}\\
K_3&=&-2\ln\left((\hr+\hr^*)/2+\ck(\hr+\hr^*)^m\hk/2^{m+2}
\right)+\fa\,,\label{K3}\\  K_4&=&-2\ln\big((\hr +\hr^*))/2
\big)-\ck(\hr+\hr^*)^{m-1}\hk/2^m+\fa\,,\label{K4}\\
K_5&=&-2\ln\left((\hr+\hr^*)/2+\ck(\hr+\hr^*)^m\hk/2^{m+2}
\right)+\nb\ln\fb\,,\label{K5}\\  K_6&=&-2\ln\big((\hr +\hr^*)/2
\big)-\ck(\hr+\hr^*)^{m-1}\hk/2^m+\nb\ln\fb\,,\label{K6} \\
K_7&=&-2\ln\left(\hr
+\hr^*\right)/2+\nb\ln\lf-\ck(\hr+\hr^*)^{m-1}\hk/2^m\nb+\fb\rg\,.\label{K7}
\eea\eeqs
Namely, $K_2$ is constructed placing $\hk$ outside the argument of
the logarithm. If we do the same for $\fa$ we can obtain two
others $K$'s, $K_3$ and $K_4$. Moreover, if we employ $\nb\ln\fb$
instead of $\fa$ to stabilize $S$ \cite{su11}, we can obtain $K_5$
and $K_6$ which have the form of $K_3$ and $K_4$ correspondingly.
Furthermore, allowing the term including $\hk$ to share the same
logarithmic argument with $\fb$ we can obtain $K=K_7$.

\subsubsection{Semi-Polynomial \Kaa\ Potentials}\label{fhi3}

Due to the exponential factor in \Eref{Vsugra},  $\Vhi$ in
\Eref{VJe} can be obtained by semi-polynomial $K$'s too -- in
contrast to \cref{nMkin} where only logarithmic $K$'s are suited.
Indeed, we seek the following
\beqs\beq K_8=-\he
-\he^*-\ck(\hr+\hr^*)^{m-1}\hk/2^m+\fa\,.\label{K8}\eeq
For $m=\ck=1$ and $\ca=b/\sqrt{2}$ [$\ca=b^2/2$], $K_8$ recovers
the $K$'s adopted in \cref{nan} for $p=1$ [$p=2$] -- no higher
order term in $F_{1S}$ is considered there. For the same $m$ and
$\ck$, $\ca=3\xi/2$ and $p=2$, $K_8$ also yields one of the $K$'s
used in \cref{lindeExp} in conjunction with a generalized version
of $W$ in \Eref{Wn} for $n=2$. Employing the alternative kinetic
terms for $S$ we can construct other two $K$'s
\bea K_9&=&-\he
-\he^*-\ck(\hr+\hr^*)^{m-1}\hk/2^m+\nb\ln\fb\,,\label{K9} \\
K_{10}&=&-\he
-\he^*+\nb\ln\lf-\ck(\hr+\hr^*)^{m-1}\hk/2^m\nb+\fb\,\rg\,,\label{K10}
\eea\eeqs
where the structure of the terms beyond $\he$ and $\he^*$ is this
adopted in Eqs.~(\ref{K6}) and (\ref{K7}).

\subsubsection{Enhanced Symmetries}\label{fhi4}

For $\rs\ll1$, our models are completely natural in the 't Hooft
sense because, in the limits $\ca\to0$ and $\ld\to0$, $K_i$ with
$i=1,...,4$ and $8$  enjoy the following enhanced symmetries:
\beqs\beq \Phi \to\ \Phi^*,\>\>\Phi \to\
\Phi+c\>\>\>\mbox{and}\>\>\> S \to\ e^{i\varphi}
S\,,\label{shift}\eeq
where $c$ and $\varphi$ are real numbers. In the same limit, $K_i$
with $i=5, 6, 7, 9$ and $10$ enjoy even more interesting enhanced
symmetries:
\beq \Phi \to\ \Phi^*,\>\>\Phi\to\ \Phi+c
\>\>\>\mbox{and}\>\>\>\frac{S}{\sqrt{\nb}}\to \frac{a
S/\sqrt{\nb}+b}{-b^*S/\sqrt{\nb}+a^*}\label{su2}\eeq\eeqs
with $|a|^2+|b|^2=1$. In other words, the theory exhibits a
$SU(2)_S/U(1)$ enhanced symmetry for the above considered $K$'s.
Thanks to the shift symmetry shown in \eqs{shift}{su2} the
operation of $\hk$ is suitably balanced as emphasized in
\crefs{nMkin, jhep, nMHkin, var}. On the one hand, its coefficient
dominates $K_{\Phi\Phi^*}$, but on the other hand, it does not
influence $\Vhi$ letting its shape intact. On the contrary,
$(\hr+\hr^{*})/2$ affects only $\Vhi$ without impact on
$K_{\Phi\Phi^*}$ (for $\ck\gg\ca$).

\subsection{The Inflaton and its Potential}\label{fhim2}

We verify below -- see \Sref{hi1} -- that $W$ and $K$'s proposed
above reproduce $\Vhi$ in \Eref{VJe}. In the region where our
models are well defined for any $p$ with $\rs\leq1$ -- see
\Sref{eff} -- $J$ in \Eref{VJe} is practically obtained too. Then,
in \Sref{hi2}, we estimate the SUGRA frame function and analyze,
in \Sref{hi3}, the stability of the inflationary path and the
possibly arising radiative corrections.

\subsubsection{Tree-Level EF Computation}\label{hi1}

Along the trough in \Eref{inftr}, the matrix with elements
$K_{\al\bbet}$ for the $K$'s in Eqs.~(\ref{K1}) -- (\ref{K10}) is
diagonal with non-vanishing elements $K_{\Phi\Phi^*}$ and
$K_{SS^*}$. In particular, we obtain
\beq \label{kpp} K_{\Phi\Phi^*}=\begin{cases} J^2,\>\>\>&\mbox{for}\>\>\>K=K_i\>\>\>\mbox{with}\>\>i=1,2\\
J^2-p^2\ca^2\sg^{2(p-1)}/2,\>\>\>&\mbox{for}\>\>\>K=K_i\>\>\>\mbox{with}\>\>i=3,...,7\\
J^2-3p^2\ca^2\sg^{2(p-1)}/2,\>\>\>&\mbox{for}\>\>\>K=K_i\>\>\>\mbox{with}\>\>i=8,9,10\end{cases}\eeq
where $J$ is given by \Eref{VJe}. These differences in the
normalization of $\se$ in the two latter cases have negligible
impact on the results during nMI for $\rs\leq1$ and any $i$ or
$i=1,...,7$ and any $\rs$. A crucial ramification is implied only
to the domains where the corresponding effective theories preserve
the perturbative unitarity up to $\mP$ for $p=1$ -- see
\Sref{eff}. Moreover, we get
\beq \label{kss} \>\>\>K_{SS^*}=\begin{cases} 1/\fr\\1\end{cases}
\mbox{for}\>\>\> K=\begin{cases} K_i&\mbox{with}\>\>i=1,2\\
K_{i}&\mbox{with}\>\>i= 3,...,10.\end{cases} \eeq
Upon substitution of $K^{SS^*}=1/K_{SS^*}$ and
\beq \label{eK}  e^K=\begin{cases} \fr^{-3}\\
\fr^{-2}\end{cases}
\mbox{for}\>\>\> K=\begin{cases} K_i&\mbox{with}\>\>i=1,2\\
K_{i}&\mbox{with}\>\>i= 3,...,10\end{cases}\eeq
into \Eref{1Vhio} we easily deduce that $\Vhi$ in \Eref{VJe} is
recovered.

A reliable extraction of the inflationary observables pre-requires
the accurate determination of the canonically normalized inflaton.
This is done inserting \eqs{kpp}{kss} in the second term of the
\emph{right-hand side} ({\sf\ftn r.h.s}) of \Eref{Saction1} which
is written as
\beqs\bea K_{\al\bbet}\dot z^\al \dot z^{*\bbet}=
{K_{\Phi\Phi^*}\over 2}\lf\dot \sg^2+\sg^2\dot\theta^2 \rg +
\frac{K_{SS^*}}{2}\lf\dot s^2+\dot{\bar
s}^2\rg\simeq\frac12\lf\dot{\widehat
\sg}^2+\dot{\what{\theta}}^{~2}+\dot{\what s}^2+{\dot{\what{\bar
s}}}^{~2}\rg, \label{Snik}\eea
where the dot denotes derivation w.r.t the cosmic time, $t$ and
the EF canonically normalized (hatted) fields can be expressed in
terms of the initial (unhatted) ones via the relations
\beq \label{VJesusy}
\frac{d\se}{d\sg}=\sqrt{K_{\Phi\Phi^*}},\>\>\widehat{\theta}
=\sqrt{K_{\Phi\Phi^*}}\sg\theta\>\>\mbox{and}\>\>(\what
s,\what{\bar s})=\sqrt{K_{SS^*}}(s,\bar s)\,.\eeq\eeqs
The spinors $\psi_\Phi$ and $\psi_S$ associated with $S$ and
$\Phi$ are normalized similarly, i.e.,
$\what\psi_{S}=\sqrt{K_{SS^*}}\psi_{S}$ and
$\what\psi_{\Phi}=\sqrt{K_{\Phi\Phi^*}}\psi_{\Phi}$, from which we
can derive the mass eigenstates
$\widehat{\psi}_{\pm}\simeq(\what\psi_{S}\pm\what\psi_{\Phi})/\sqrt{2}$
-- see below.

\subsubsection{Frame Function}\label{hi3}

Limiting ourselves along the inflationary trajectory in
\Eref{inftr} we can define the function $\Omega$ via the relation
\beq\label{omgdef} {\Omega/N}=-e^{-K/N},~~\mbox{where}~~
N=\begin{cases}
3,\>\>\>\mbox{for}\>\>\>K=K_i\>\>\>\mbox{with}\>\>\>i=1,2
\\2,\>\>\> \mbox{for}\>\>\>K=K_i\>\>\>\mbox{with}\>\>\>i=3,...,10.\end{cases}
\eeq
If we perform the inverse of the conformal transformation
described in \eqs{action1}{action} along the lines of \cref{jhep}
we end up with the JF potential $\Vjhi=\Omega^2\Vhi/N^2$ in
\Eref{Vn} with the function $-\Omega/N=\fr$ acting as a frame
function. Moreover, the conventional Einstein gravity at the SUSY
vacuum, $\vev{S}=\vev{\Phi}=0$, is recovered since
$-\vev{\Omega}/N=1$.

\subsubsection{Stability and one-Loop Radiative
Corrections}\label{hi2}

Contrary to the non-SUSY case where the inflaton appears uncoupled
in \Eref{action}, here it coexists obligatorily with other scalars
and fermions which construct the two chiral supermultiplets -- for
possible ramifications of the non-SUSY nMI due to the presence of
couplings between the inflaton with other fields see \cref{quad}.
Therefore, we have to verify that the inflationary direction in
\Eref{inftr} is stable w.r.t the fluctuations of the non-inflaton
fields. To this end, we construct the mass-spectrum of the scalars
taking into account the canonical normalization of the various
fields in \Eref{Snik}. In the limit $\fr>1$, we find the
expressions of the masses squared $\what m^2_{\chi^\al}$ (with
$\chi^\al=\theta$ and $s$) arranged in Table~1. These results
approach rather well the quite lengthy, exact expressions taken
into account in our numerical computation and are valid for any
$\rs$. From these findings we can easily confirm that $\what
m^2_{\chi^\al}\gg\Hhi^2=\Vhi/3$ during nMI provided that
$\kx>1/6\fr$ for $K=K_i$ with $i=1$ and $2$ or $\kx>0.2$ for
$K=K_i$ with $i=3,4$ and $8$ or $0<\nb<6$ for $K=K_i$ with $i=5,6,
7, 9$ and $10$. This means that the fields $\chi^\al$ rapidly roll
towards zero and stay there during nMI.

\setcounter{table}{1}
\renewcommand{\arraystretch}{1.4}
\begin{sidewaystable}[h!]
\vspace*{-15.0cm} \bec
\begin{tabular}{|c|c|c||c|c|c|c|c|c|c|c|}\hline
{\sc Fields}&{\sc Eigen-} & \multicolumn{9}{c|}{\sc Masses
Squared}\\\cline{3-11}
&{\sc states}&& {$K=K_1$}&{$K=K_2$} &{$K=K_{j+2}$}&$K=K_{j+4}$&$K=K_7$&$K=K_8$&$K=K_9$&$K=K_{10}$    \\
&&& &&{$(j=1,2)$}&$(j=1,2)$&&&&    \\
\hline\hline
%
%
1 {real scalar}&$\widehat\theta_{+}$&$\widehat
m_{\theta+}^2/\Hhi^2$&
$4$&\multicolumn{3}{|c|}{$6$}&$6(1+1/\nb)$&\multicolumn{2}{|c|}{$6$}&$6(1+1/\nb)$\\\cline{3-11}
1 complex scalar&$\widehat s, \widehat{\bar{s}}$ & $ \widehat m_{
s}^2/\Hhi^2$&\multicolumn{2}{c|}{$2\lf6\kx\fr-1\rg$}&{$12\kx$}&\multicolumn{2}{|c|}{$6/\nb$}&$12\kx$&
\multicolumn{2}{|c|}{$6/\nb$}\\\cline{3-11}
$2$ Weyl spinors & $\what \psi_\pm $ & $\what m^2_{
\psi\pm}/\Hhi^2$ & \multicolumn{8}{c|}{${3(2p\ca
\sg^p-n)^2}/{2\ck\sg^2\fr^{m-1}}$}\\\hline
\end{tabular}\\[0.5cm]
{\slshape\bfseries \small Table~1:} {\sl\small Mass-squared
spectrum for $K=K_i$ with $i=1,...,10$ along the inflationary
trajectory in \Eref{inftr} for $\sg\ll1$.}
\end{center}
\end{sidewaystable}

\clearpage

Inserting the derived mass spectrum in the well-known
Coleman-Weinberg formula \cite{cw}, we can find the one-loop
radiative corrections, $\dV$ to $\Vhi$. It can be verified that
our results are immune from $\dV$, provided that the
renormalization group mass scale $\Lambda$ -- involved in this
formula --, is determined by requiring $\dV(\sgx)=0$ or
$\dV(\sgf)=0$. The possible dependence of our results on the
choice of $\Lambda$ can be totally avoided if we confine ourselves
to $\kx\sim(0.5-1.5)$ in $K_i$ with $i=1,...,4$ or $0<\nb<6$ in
$K_i$ with $i=5,6$ and $7$ resulting to
$\Ld\simeq(0.9-9)\cdot10^{-5}$ -- cf. \crefs{quad,jhep}. Under
these circumstances, our results can be reproduced by using
exclusively $\Vhi$ in \Eref{VJe}.

\section{Inflation Analysis}\label{fhi}

A successful inflationary scenario has to be compatible with a
number of criteria  which are outlined in Sec.~\ref{const}. We
then test our models against these constraints, first numerically
in Sec.~\ref{num} and then (semi)analytically in Sec.~\ref{resa}.

\subsection{Inflationary Constraints} \label{const}

The analysis of nMI can be carried out exclusively in the EF using
the standard slow-roll approximation keeping in mind the
dependence of $\what\phi$ on $\phi$ -- given by \Eref{VJe} in both
the SUSY and non-SUSY set-up. Working this way, we outline in the
following a number of observational and theoretical requirements
which we impose in our investigation -- see, e.g., \cref{review1,
review}.

\subsubsection{Inflationary {\rm e}-Foldings} The number of
e-foldings that the pivot scale $\ks=0.05/{\rm Mpc}$ experiences
during nMI, has to be enough to resolve the horizon and flatness
problems of standard big bang cosmology, i.e., \cite{plin,nmi}
\begin{equation}
\label{Nhi}  \Ns=\int_{\se_{\rm f}}^{\se_\star}\, d\se\:
\frac{\Ve_{\rm CI}}{\Ve_{\rm CI,\se}}
\simeq61.3+\frac12\ln\lf{\Vhi(\sgx)\over\Vhi(\sgf)^{1/2}}{\fr(\sgx)\over
g_{\rm rh*}^{1/6}}\rg + \frac{1-3w_{\rm rh}}{12(1+w_{\rm rh})}\ln
\frac{\pi^2g_{\rm rh*}\Trh^4}{30\Vhi(\sgf)\fr(\sgf)^2}
\end{equation}
where we assumed that nMI is followed in turn by a oscillatory
phase, with mean equation-of-state parameter $w_{\rm rh}$
\cite{turner}, radiation and matter domination. Also $\Trh$ is the
reheat temperature after nMI, with energy-density effective number
of degrees of freedom $g_{\rm rh*}=106.75$ which corresponds to
the Standard Model spectrum. Since, for low values of $\phi$, our
inflationary potentials can be well approximated by a $\phi^n$
potential we compute $w_{\rm rh}$ \cite{turner,review} from the
formula
\begin{equation}
\label{wrh} w_{\rm rh} = \frac{n-2}{n+2}\>\>\>\Rightarrow\>\>\> w_{\rm rh}=\begin{cases}0,&\mbox{for}\>\>n=2\\
1/3,&\mbox{for}\>\>n=4.\end{cases}
\end{equation}
Note that for $n=4$, $\Ns$ turns out to be independent of $\Trh$
-- cf. \crefs{var,jhep}. For $n=2$ we take
$\Trh=4.1\cdot10^{-10}$. In \Eref{Nhi} $\sgx~[\sex]$ is the value
of $\sg~[\se]$ when $\ks$ crosses outside the inflationary
horizon, and $\sgf~[\sef]$ is the value of $\sg~[\se]$ at the end
of nMI, which can be found, in the slow-roll approximation, from
the condition
\beq{\ftn\sf
max}\{\eph(\se),|\ith(\se)|\}\simeq1,\>\mbox{where}\>\>
\eph=\frac12\left(\frac{\Ve_{\rm CI,\se}}{\Ve_{\rm
CI}}\right)^2\>\>\>\mbox{and}\>\> \ith={\Ve_{\rm
CI,\se\se}\over\Ve_{\rm CI}}\cdot \label{srcon} \eeq

\subsubsection{Normalization of the Power Spectrum} The amplitude $\As$ of the
power spectrum of the curvature perturbation generated by $\sg$ at
the pivot scale $k_\star$ must to be consistent with
data~\cite{plcp}
\begin{equation}  \label{Prob}
\sqrt{A_{\rm s}}=\: \frac{1}{2\sqrt{3}\, \pi} \; \frac{\Ve_{\rm
 CI}(\sex)^{3/2}}{|\Ve_{\rm
 CI,\se}(\sex)|}=\frac{|J(\sgx)|}{2\sqrt{3}\, \pi} \;
\frac{\Ve_{\rm CI}(\sgx)^{3/2}}{|\Ve_{\rm
 CI,\sg}(\sgx)|}\simeq4.627\cdot 10^{-5},
\end{equation}
where we assume that no other contributions to the observed
curvature perturbation exists. This is ensured from the heaviness
of the non-inflaton fields checked in \Sref{hi2}.

\subsubsection{Inflationary Observables}\label{nspl}

The remaining inflationary observables (the spectral index $\ns$,
its running $\as$, and the tensor-to-scalar ratio $r$) must be in
agreement with the fitting of the \plk, \emph{Baryon Acoustic
Oscillations} ({\sf\ftn BAO}) and \bcp\ data \cite{plin,gwsnew}
with $\Lambda$CDM$+r$ model, i.e.,
\begin{equation}  \label{nswmap}
\mbox{\ftn\sf
(a)}\>\>\ns=0.968\pm0.009\>\>\>~\mbox{and}\>\>\>~\mbox{\ftn\sf
(b)}\>\>r\leq0.07,
\end{equation}
at 95$\%$ \emph{confidence level} ({\sf\ftn c.l.}) with
$|\as|\ll0.01$. Although compatible with \sEref{nswmap}{b} all
data taken by the \bcp\ CMB polarization experiments up to and
including the 2014 observing season ({\sf\ftn BK14}) \cite{gwsnew}
seem to favor $r$'s of order $0.01$ since $r=
0.028^{+0.025}_{-0.025}$ at 68$\%$ c.l. has been reported. These
inflationary observables are estimated through the relations:
\beq\label{ns} \mbox{\ftn\sf (a)}\>\>\> \ns=\: 1-6\eph_\star\ +\
2\ith_\star,\>\>\>\mbox{\ftn\sf (b)}\>\>\>
\as=\:\frac23\left(4\widehat\eta_\star^2-(\ns-1)^2\right)-2\widehat\xi_\star\>\>\>\mbox{and}\>\>\>\mbox{\ftn\sf
(c)}\>\>\>r=16\eph_\star, \eeq
where $\widehat\xi={\Ve_{\rm CI,\se} \Ve_{\rm
CI,\se\se\se}/\Ve_{\rm CI}^2}$ and the variables with subscript
$\star$ are evaluated at $\sg=\sgx$. For a direct comparison of
our findings with the obervational outputs in \crefs{plin,gwsnew},
we also compute $\rw=16\eph(\se_{0.002})$ where $\se_{0.002}$ is
the value of $\se$ when the scale $k=0.002/{\rm Mpc}$, which
undergoes $\what N_{0.002}=\Ns+3.22$ e-foldings during nMI,
crosses outside the horizon of nMI.

\subsubsection{Tuning of the Initial Conditions}

In all cases $\Vhi$ in \Eref{VJe} develops a local maximum
\beq \Vhi(\sg_{\rm max})=2^{-n(2+p)/2p}\ld^2
\lf\frac{ep\ca}{n}\rg^{n/p}\>\>\>\mbox{at}\>\>\> \sg_{\rm
max}=\frac1{\sqrt{\ck}}\lf\frac{n}{2p\rs}\rg^{1/p}\,,\label{Vmax}\eeq
giving rise to a stage of hilltop \cite{lofti} nMI. In a such case
we are forced to assume that \FHI occurs with $\sg$ rolling from
the region of the maximum down to smaller values. Therefore, a
mild tuning of the initial conditions is required which can be
quantified somehow defining \cite{gpp} the quantity:
\beq \Dex=\left(\sg_{\rm max} - \sgx\right)/\sg_{\rm
max}\,.\label{dms}\eeq
The naturalness of the attainment of \FHI increases with $\Dex$
and it is maximized when $\sg_{\rm max}\gg\sgx$ or $\Dex\simeq1$.
This is facilitated as $\rs$ and $p$ decrease and $n$ increases.

\subsubsection{Effective Field Theory} \label{eff}

Our inflationary scenario becomes stable against corrections of
higher order non-renormalizable terms if we impose two additional
theoretical constraints -- keeping in mind that
$\Vhi(\sgf)\leq\Vhi(\sgx)$:
\beq \label{subP}\mbox{\ftn\sf (a)}\>\> \Vhi(\sgx)^{1/4}\leq\Qef
\>\>\>\mbox{and}\>\>\>\mbox{\ftn\sf (b)}\>\> \sgx\leq\Qef\,,\eeq
where $\Qef$ is the \emph{ultaviolet} ({\ftn\sf UV}) cutoff scale.
Below we show that, in the non-SUSY regime, $\Qef=1$ (in units of
$\mP$) for $p=1$ and any $\rs$ or $p>1$ and $\rs\leq1$. As a
consequence, no concerns regarding the validity of the effective
theory arise although $\ck$ (or $\ca$) may take relatively large
values for $\sg<1$ -- see Secs.~\ref{num} and \ref{resa}. The
origin of this nice behavior is the fact that the EF (canonically
normalized) inflaton,
\beq\dphi=\vev{J}\dph\>\>\>\mbox{with}\>\>\> \dph=\phi-\vev{\sg}
\>\>\>\mbox{and}\>\>\>\vev{J}=\begin{cases}
\sqrt{\ck+3\ca^2/2},&\mbox{for}\>\>\>p=1
\\ \sqrt{\ck},&\mbox{for}\>\>\>p>1\end{cases}
\label{dphi} \eeq does not coincide with $\dph$ at the vacuum of
the theory -- contrary to the pure nMI \cite{cutoff, riotto} with
$n>2$ -- for $\ck\gg1$ and any $p$ or even for $\ck\ll1, \ca\gg1$
and $p=1$.

To clarify further this point we analyze the small-field behavior
of our models in the EF. We restrict ourselves to the $(n, p)$'s
which support acceptable inflationary solutions, shown in
\Sref{num}. Although the expansions about $\vev{\phi} = 0$,
presented below, are not valid during nMI, we consider $\Qef$
extracted this way as the overall cut-off scale of the theory,
since the reheating phase – realized via oscillations about
$\vev{\phi}=0$ is an unavoidable stage of the inflationary
dynamics. Namely, we focus on the second term in the r.h.s of
\Eref{action} for $\mu=\nu=0$ and we expand it about
$\vev{\phi}=0$ in terms of $\se$. Our result can be written as
\beqs\beq\label{Jexp}  J^2 \dot\phi^2\simeq\dot\se^2\cdot
\begin{cases} 1 + \frac{2 \sqrt{2}(m-1)\rs\se}{(2 + 3\rs^2)^{3/2}}
+ \frac{2 (m-1)^2\rs^2\se^2}{(2 + 3\rs^2)^2} + \frac{2 \sqrt{2}
(m-1)^3\rs^3\se^3}{3 (2 + 3\rs^2)^{5/2}},&\mbox{for}\>\>\>p=1
\\ 1+(m-1)\rs\se^2 + (m-1)^2\rs^2\se^4/2 ,&\mbox{for}\>\>\>p=2\,.\end{cases}\eeq
The form of the expansions above for $p=1$ can be specified in the
two following regimes
\beq\label{Jexp1}  J^2 \dot\phi^2\simeq\dot\se^2\cdot
\begin{cases} 1 +(m-1)\rs\se + (m-1)^2\rs^2\se^2/2 +
(m-1)^3\rs^3\se^3/6,&\mbox{for}\>\>\>\rs\ll 1
\\ 1 +\frac1{\rs^2}\lf \frac23\sqrt{\frac23} (m-1)\se - \frac2{9}(m-1)^2\se^2
+\frac{2}{27}\sqrt{\frac23}(m-1)^3\se^3\rg,&\mbox{for}\>\>\>\rs\gg1\,.\end{cases}\eeq
We remark that no $\rs$ appears in the denominators for $\rs\ll1$
and no $\rs$ appears in the nominators for $\rs\gg1$ and so
$\Qef=1$. Note also that as $\rs$ approaches unity, $m-1$ has not
to exceed a lot unity. We impose conservatively a global bound
$m\leq15$ -- see \Sref{num}.

Expanding similarly $\Vhi$, see \Eref{VJe}, in terms of $\se$ we
have
\beq \Vhi\simeq\frac{\ld^2\se^n}{(2\ck)^{n/2}}\cdot\begin{cases}
1 - 2\rs\se + 2\rs^2\se^2 - 4\rs^3\se^3/3 + 2\rs^4\se^4/3,&\mbox{for}\>\>\>(n,p)=(2,1) \\
1 - 2\rs\se^2 + 2\rs^2\se^4-4\rs^6\se^3/3,&\mbox{for}\>\>\>(n,p)=(2,2) \\
1 - 2\rs\se + 2\rs^2\se^2 - 4\rs^3\se^3/3 +
(2\rs^4\se^4)/3,&\mbox{for}\>\>\>(n,p)=(4,1)
\\ 1 - 2\rs\se^2 + 2\rs^2\se^4-4\rs^6\se^3/3,&\mbox{for}\>\>\>(n,p)=(4,2)\,.\end{cases}
\label{Vexp}\eeq
For $p=2$ the expansions above are valid for any $\rs$ whereas for
$p=1$ these are convenient only for $\rs\ll1$. For $\rs\gg1$ and
$p=1$, $\Vhi$ may be expanded as follows
\beq \Vhi\simeq\frac{2\ld^2\se^n}{(2\ck)^{n/2}}\cdot\begin{cases}
(27 - 18 \sqrt{6}\se + 36\se^2- 8 \sqrt{6}\se^3+8\se^4)/81\rs^2+{\cal O}(\rs^{-4}),&\mbox{for}\>\>\>(n,p)=(2,1) \\
2(27 - 18 \sqrt{6}\se + 36\se^2- 8 \sqrt{6}\se^3)/243\rs^4+{\cal
O}(\rs^{-5}),&\mbox{for}\>\>\>(n,p)=(4,1).\end{cases}
\label{Vexp1}\eeq\eeqs
We conclude again that $\Qef=1$ independently from $m$ for the
ranges of $p$ and $\rs$ mentioned below \Eref{subP} -- note that
the expressions above for $p=2$ can be easily extended to all
$p>1$.

Taking into account \Eref{kpp} we infer that the results above are
also valid for our SUGRA models with $p>1$ since
$\vev{K_{\Phi\Phi^*}}=\vev{J^2}$. When $p=1$, we obtain
$\vev{K_{\Phi\Phi^*}}=\vev{J^2}=\ck+3\ca/2$ for $K=K_i$ with
$i=1,2$, $\vev{K_{\Phi\Phi^*}}=\ck+\ca$ for $K=K_i$ with
$i=3,...,7$ and $\vev{K_{\Phi\Phi^*}}=\ck$ for $K=K_i$ with
$i=8,9,10$. Therefore, for $p=1$ and $i=1,2$, the expansions of
$K_{\Phi\Phi^*}\dot{\sg}^2$ and $\Vhi$ coincide with those given
for $J^2\dot{\sg}^2$ and $\Vhi$ above. For $p=1$ and $i=3,...,7$,
expansions similar to the ones obtained in the non-SUSY case can
be extracted, leading to identical conclusions. On the contrary,
for $p=1$ and $i=8,9,10$ the expansions for any $\rs$ coincide
with the expansions above for $\rs\ll1$. Therefore, in the last
case, naturalness implies $\rs\leq1$. In other words, for $p=1$
there is a theoretical discrimination between our SUGRA models
based on the (semi)logarithmic and semipolynomial $K$'s.

Our overall conclusion is that our models respect the perturbative
unitarity up to $\mP$ for $\rs\leq1$ and $p>1$. For $p=1$ the
non-SUSY models and the SUGRA ones for $K=K_i$ with $i=1,...,7$
are unitarity safe up to $\mP$ for any $\rs$, whereas the SUGRA
models relied on $K=K_i$ with $i=8,9,$ or $10$ require still
$\rs\leq1$. On the other hand, $m$ has to be of order unity for
$\rs$ approaching $1$.


\subsection{Numerical Results}\label{num}

The SUSY version of our models, which employs $W$ in \Eref{Wn} and
one of the $K$'s in Eqs.~(\ref{K1}) -- (\ref{K10}), is described
by the following parameters:
\beq \label{para}
\ld,\>n,\>p,\>m,\>\ca,\>\ck\>\mbox{and}\>\>\kx\>\mbox{or}\>\nb\,,\eeq
for the $K$'s given by Eqs.~(\ref{K1}) -- (\ref{K4}) and
(\ref{K8}) or Eqs.~(\ref{K5}) -- (\ref{K7}), (\ref{K9}) and
(\ref{K10}), respectively. Obviously, the non-SUSY models do not
depend on the two last parameters which control only $\what
m_{s}^2$ in Table~1 and let intact the inflationary predictions
provided that these are selected so that $\what m_{s}^2>\Hhi^2$.
Recall that we use $\Trh=4.1\cdot10^{-9}$ throughout and $\Ns$ is
computed self-consistently with $n$ via \eqs{wrh}{Nhi}. Our result
is $\Ns\simeq(50-52)$ for $n=2$ and $\Ns=(55-58)$ for $n=4$. Note
finally that for fixed $n, p$ and $m$, $J$ and $\Vhi$ in
\Eref{VJe} are functions of $\rs=\ca/\ck^{p/2}$ and
$\ld/\ck^{n/4}$ and not $\ca$, $\ck$ and $\ld$ as naively expected
-- see \Sref{nonsusy}.

The confrontation of the parameters above with observations is
implemented numerically substituting $J$ and $\Vhi$ in
Eqs.~(\ref{Nhi}), (\ref{srcon}), and (\ref{Prob}), and extracting
the inflationary observables as functions of $n, p, m, \rs,
\ld/\ck^{n/4}$ and $\sgx$. The two latter parameters can be
determined by fulfilling Eqs.~(\ref{Nhi}) and (\ref{Prob}). We
then compute the predictions of the models applying \Eref{ns} for
every selected $n, p, m$ and $\rs$, taking into account the
available data in \Eref{nswmap}. From \eqs{VJe}{kpp} we see that
the obtained results are precisely valid for the non-SUSY models
and the SUSY ones for $K=K_i$ with $i=1,2$. However, as emphasized
in \Sref{fhim2}, these are practically identical for any $i$ and
the $\rs$'s which assure the validity of the corresponding
effective theories up to $\mP$ -- see \Sref{eff}.



\begin{figure}[!t]\vspace*{-.in}
\hspace*{-.19in}
\begin{minipage}{8in}
\epsfig{file=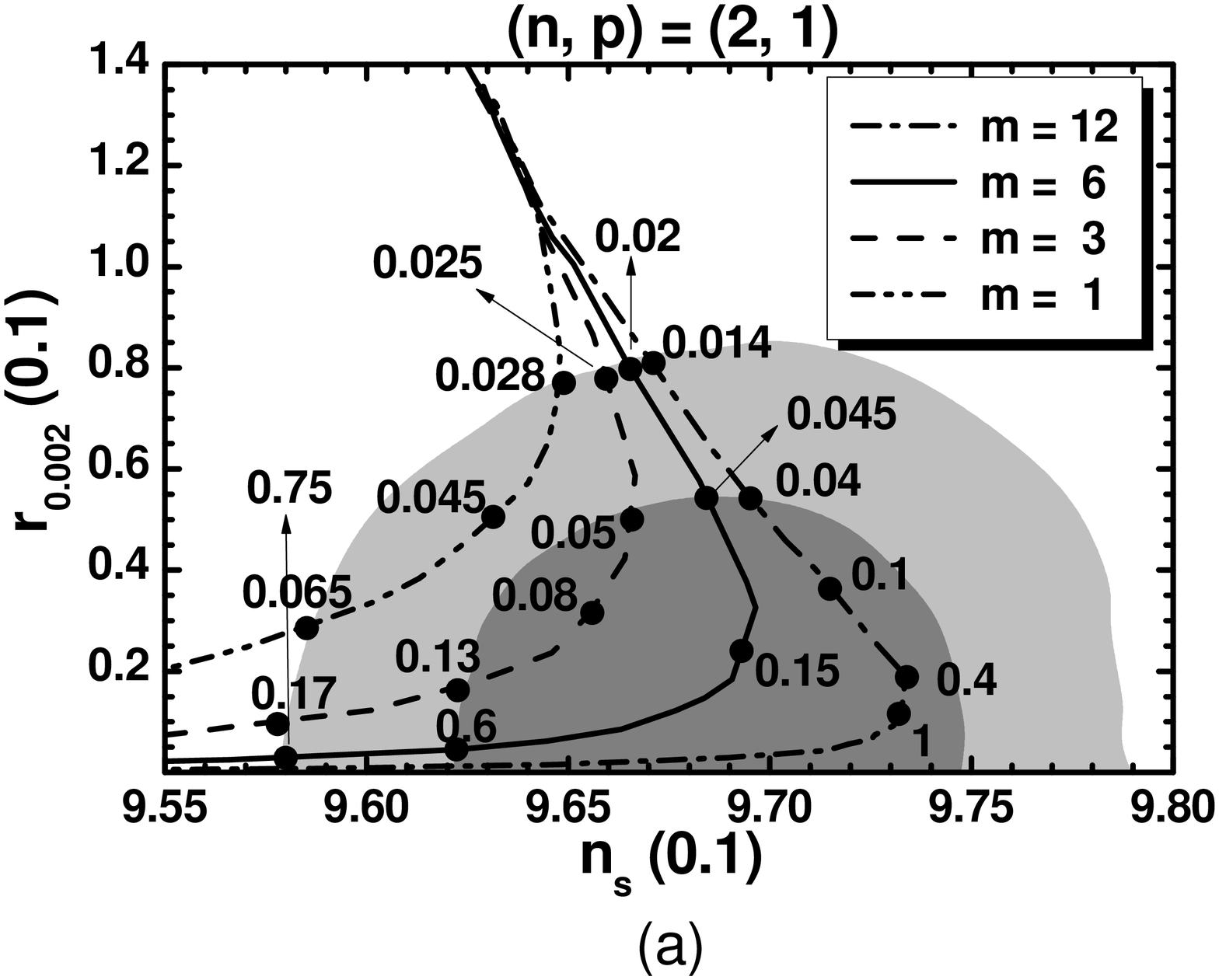,height=3.6in,angle=-90}
\hspace*{-1.2cm}
\epsfig{file=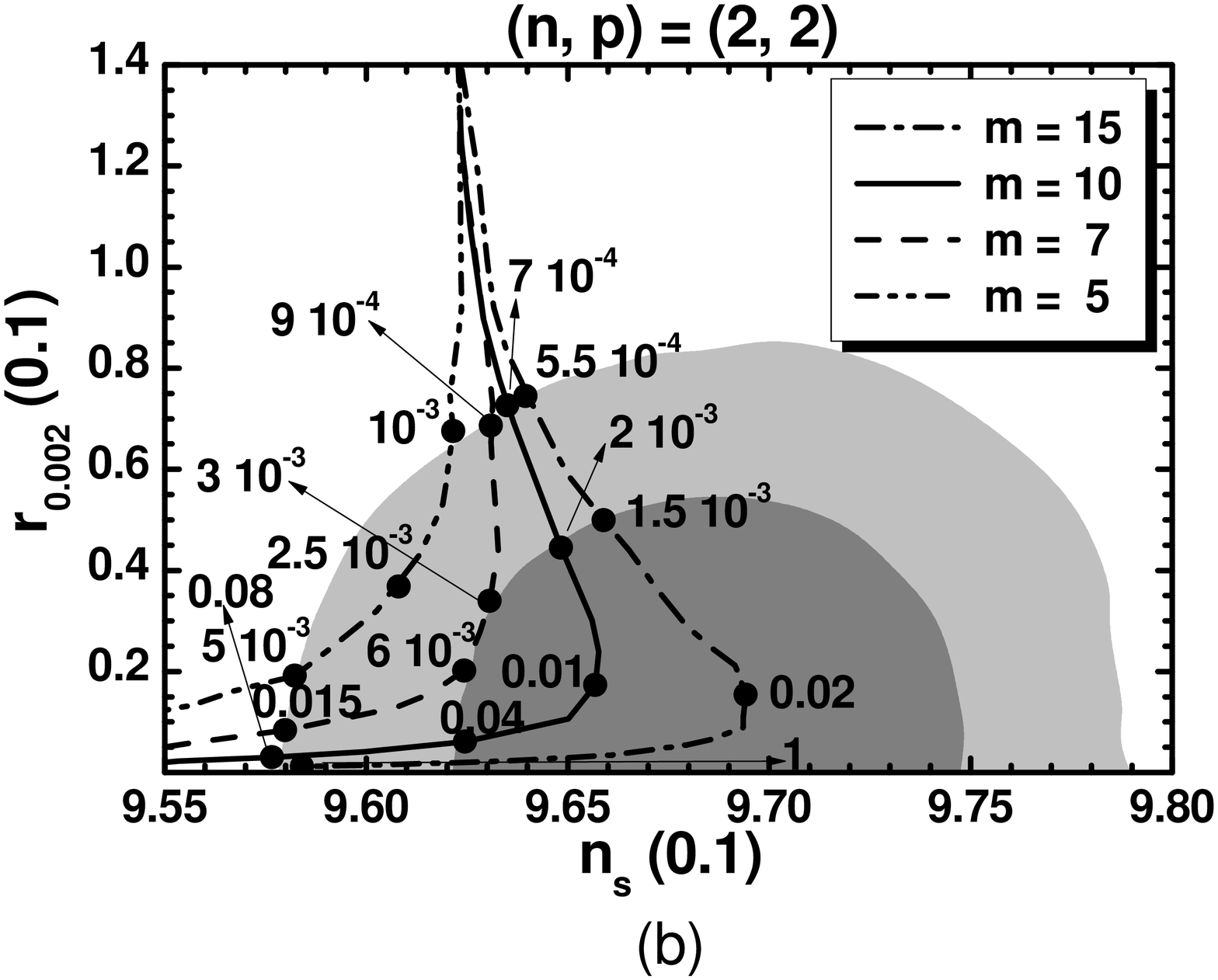,height=3.6in,angle=-90} \hfill
\end{minipage}\vspace*{-.in}
\hfill \hspace*{-.19in}
\begin{minipage}{8in}
\epsfig{file=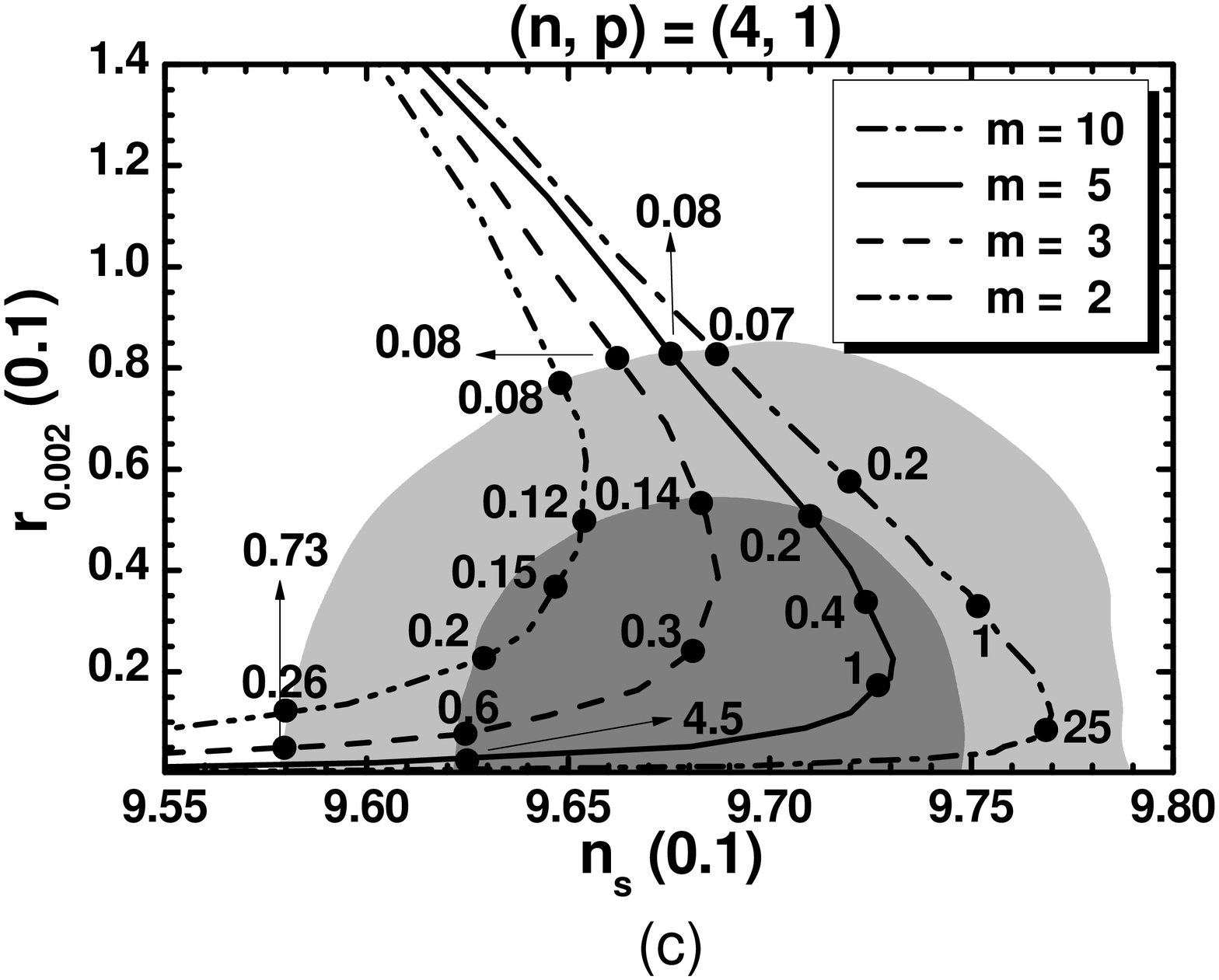,height=3.6in,angle=-90}
\hspace*{-1.2cm}
\epsfig{file=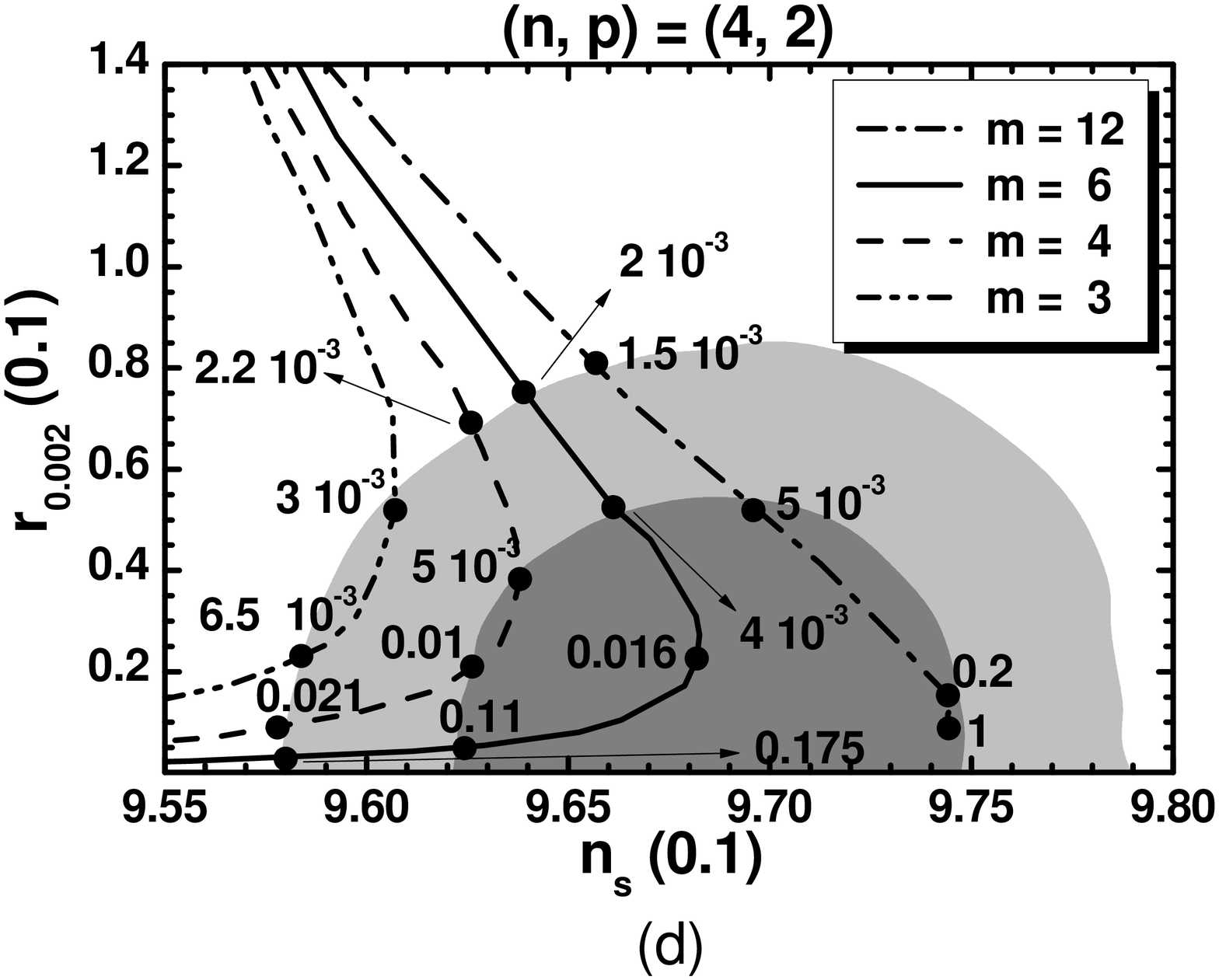,height=3.6in,angle=-90} \hfill
\end{minipage}\hfill\begin{center}\renewcommand{\arraystretch}{1.1}
{\ftn\begin{tabular}{|c||cccc|cccc|}\hline {\sc Plot}
&\multicolumn{4}{c|}{{\ssz \sffamily (a)}: $n=2$ \&
$p=1$}&\multicolumn{4}{c|}{{\ssz \sffamily (b)}: $n=2$ \&
$p=2$}\\\cline{2-9}
$m$&$1$&$3$&$6$&$12$&$5$&$7$&$10$&$15$\\\hline\hline
$\rs^{\rm min}/10^{-2}$&$2.8$&$2.5$&$2$&$1.4$&$0.1$&$0.09$&$0.07$&$0.055$\\
$\rs^{\rm max}/0.1$&$0.65$&$1.7$&$7.5$&$56~\{10\}$&$0.05$&$0.15$&$0.8$&$10$\\
$\rw^{\rm
min}/0.01$&$2.8$&$0.95$&$0.29$&$0.078~\{1.9\}$&$1.9$&$0.84$&$0.3$&$0.12$
\\\hline\hline
{\sc Plot} &\multicolumn{4}{c|}{{\ssz \sffamily (c)}: $n=4$ \&
$p=1$}&\multicolumn{4}{c|}{{\ssz \sffamily (d)}: $n=4$ \&
$p=2$}\\\cline{2-9}
$m$&$2$&$3$&$5$&$10$&$3$&$4$&$6$&$12$\\\hline\hline
$\rs^{\rm min}/10^{-2}$&$8$&$8$&$8$&$7$&$0.3$&$0.22$&$0.2$&$0.15$\\
$\rs^{\rm max}/0.1$&$2.6$&$7.3$&$55~\{10\}$&$8\cdot10^{3}~\{10\}$&$0.065$&$0.21$&$1.75$&$10$\\
$\rw^{\rm
min}/0.01$&$1.2$&$0.49$&$0.012$&$3.8\cdot10^{-3}$&$2.3$&$0.89$&$0.27$&$0.88$\\
&&&$\{1.73\}$&$\{3.3\}$&&&&
\\\hline
\end{tabular}}
\end{center}\vspace*{-.0in}\renewcommand{\arraystretch}{1.}
\hfill \vchcaption{\slshape\small  Allowed curves in the $\ns-\rw$
plane for {\sffamily\ssz (a)} $n=2$, $p=1$ and $m=1, 3, 6,$ and
$12$, {\ssz\sffamily (b)} $n=2$, $p=2$ and $m=5,7,10$ and $15$,
{\ssz\sffamily (c)} $n=4$, $p=1$ and $m=2,3,5$ and $10$,
{\sffamily\ssz (d)} $n=4$, $p=2$ and $m=3,4,6$ and $12$ with the
$\rs$ values indicated on the curves. The conventions adopted for
the various lines are shown in the plots. The marginalized joint
$68\%$ [$95\%$] regions from \plk, BAO and BK14 data are depicted
by the dark [light] shaded areas.  The allowed $\rs^{\rm min}$,
$\rs^{\rm max}$ and $\rw^{\rm min}$ values in each plot are listed
in the table. The values in curly brackets correspond to the SUSY
models with $K=K_i$ where $i=8,9$ and $10$. In these cases the
allowed curves are limited to values $\rs\leq1$. }\label{fig1}
\end{figure}\vfill

We start the presentation of our results by comparing the outputs
of our models against the observational data \cite{plin,gwsnew} in
the $\ns-\rw$ plane -- see \Fref{fig1}. We depict the
theoretically allowed values for {\ftn\sf (i)} $(n,p)=(2,1)$ and
$m=1,3,6$ and $12$ in \sFref{fig1}{a}, {\ftn\sf (ii)}
$(n,p)=(2,2)$ and $m=5,7,10$ and $15$ in \sFref{fig1}{b}, {\ftn\sf
(iii)} $(n,p)=(4,1)$ and $m=2,3,5$ and $10$ in \sFref{fig1}{c},
and {\ftn\sf (iv)} $(n,p)=(4,2)$ and $m=3,4,6$ and $12$ in
\sFref{fig1}{d}. The conventions adopted for the various lines are
shown in the plots and the variation of $\rs$ is shown along each
of them. In particular, we use double dot-dashed, dashed, solid
and dot-dashed lines for each $(n,p)$ with increasing $m$. For
$m=1$ we confirm the findings of \cref{nan} according to which the
lines move to the left increasing $n$ with fixed $p$. Indeed, for
$n=2$ and $p=m=1$ the double dot-dashed curve lies inside the
observationally favored (light gray) region as seen from
\sFref{fig1}{a}. For $n=4$ and $p=m=1$, however, the corresponding
line lies entirely outside (and to the left of) the allowed
region. For this reason we plot the line with $m=2$ which displays
an observationally acceptable segment as shown in \sFref{fig1}{c}.
On the contrary, the lines with the same $p$ and $m>1$ have the
tendency to move to the right for increasing $n$ as can be
inferred comparing the position of dashed lines in \sFref{fig1}{a}
and {\sf\ftn (c)}. This fact indicates that the kinetic mixing
introduced in \Eref{fk} plays a key role for the viability of our
models.

In all plots of \Fref{fig1}, we observe that for low enough $\rs$
-- i.e. $\rs\leq0.0001$ -- the various lines converge to the
$(\ns,\rw)$'s obtained within the minimal chaotic inflation
defined for $\ca=0$, i.e., $(\ns,\rw)\simeq(0.962,0.14)$ in
\sFref{fig1}{a} and {\sf\ftn (b)} and
$(\ns,\rw)\simeq(0.949,0.25)$ (not shown in the plots) in
\sFref{fig1}{c} and {\sf\ftn (d)}. Increasing $\rs$ we can
determine a minimal $\rs$, $\rs^{\rm min}$, for which the various
lines enter the $95\%$ c.l. observationally allowed region. For
$\rs>\rs^{\rm min}$ the various lines cover the marginalized joint
$95\%$ c.l. regions, turn to the left and mostly cross outside
them.  On each line we can also define a maximal $\rs$, $\rs^{\rm
max}$, which obviously correspond to a minimal $\rw$, $\rw^{\rm
min}$. We have $\rs^{\rm max}=1$, if the theory ceases to be
unitarity safe beyond this value, as for $(n, p, m)=(2,2,15)$ and
$(4,2,12)$. Otherwise, $\rs^{\rm max}$ is the $\rs$ value for
which the corresponding lines cross outside the $95\%$ c.l.
observationally allowed corridors. More specifically, the
$\rs^{\rm min}$, $\rs^{\rm max}$ and $\rw^{\rm min}$ values for
each line are accumulated in the table shown below the plots. Note
that, if we employ $K=K_i$ with $i=8,9$ or $10$ for the SUSY
implementation of our models, the dot-dashed line in
\sFref{fig1}{a} and the solid and dot-dashed lines in
\sFref{fig1}{c} have to be terminated at $\rs^{\rm max}=1$ and the
relevant $\rs^{\rm max}$ and $\rw^{\rm min}$ values are displayed
in curly brackets.

\begin{figure}[!t]
\hspace*{-.19in}
\begin{minipage}{8in}
\epsfig{file=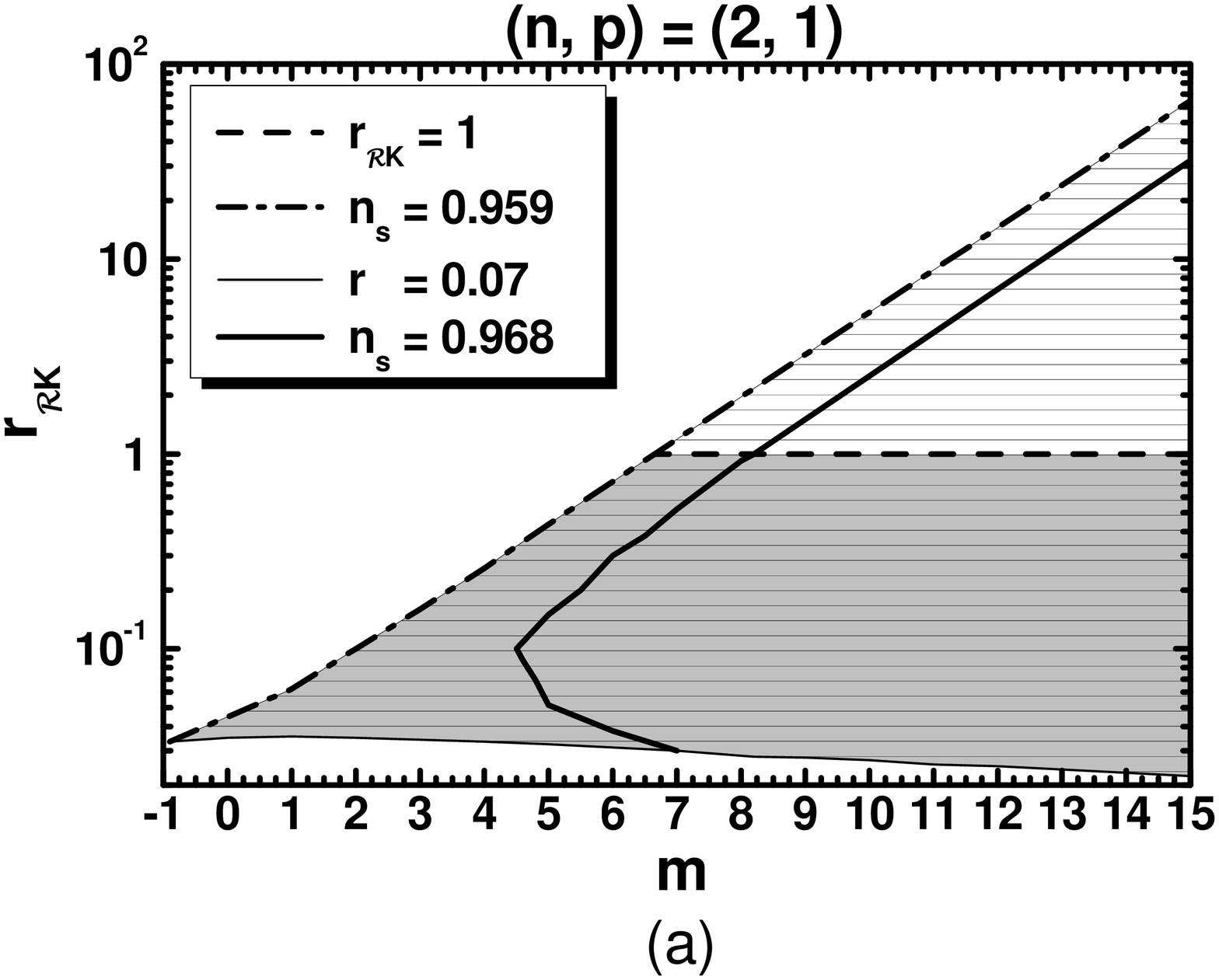,height=3.6in,angle=-90}
\hspace*{-1.2cm}
\epsfig{file=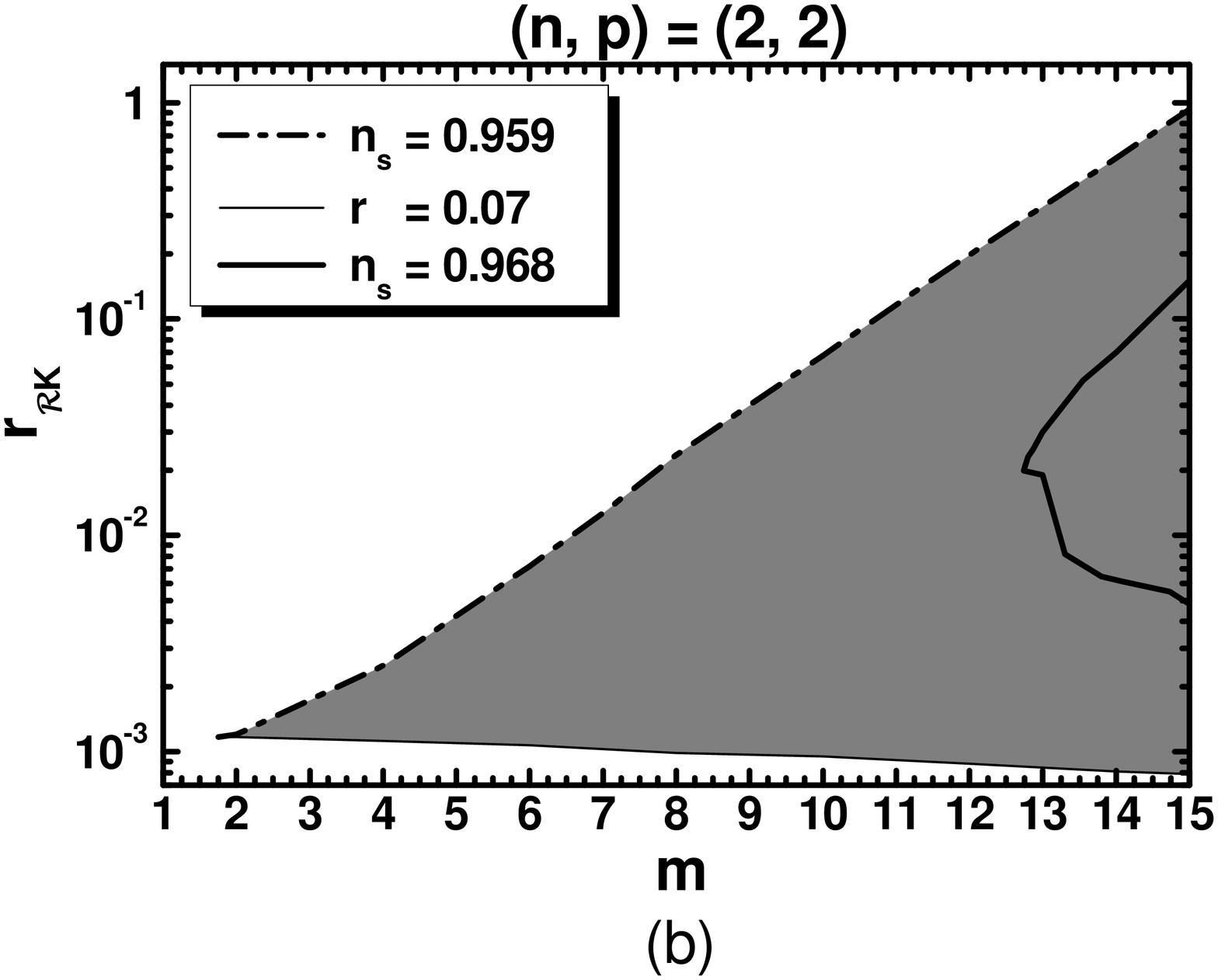,height=3.6in,angle=-90} \hfill
\end{minipage}
\hfill \hspace*{-.19in}
\begin{minipage}{8in}
\epsfig{file=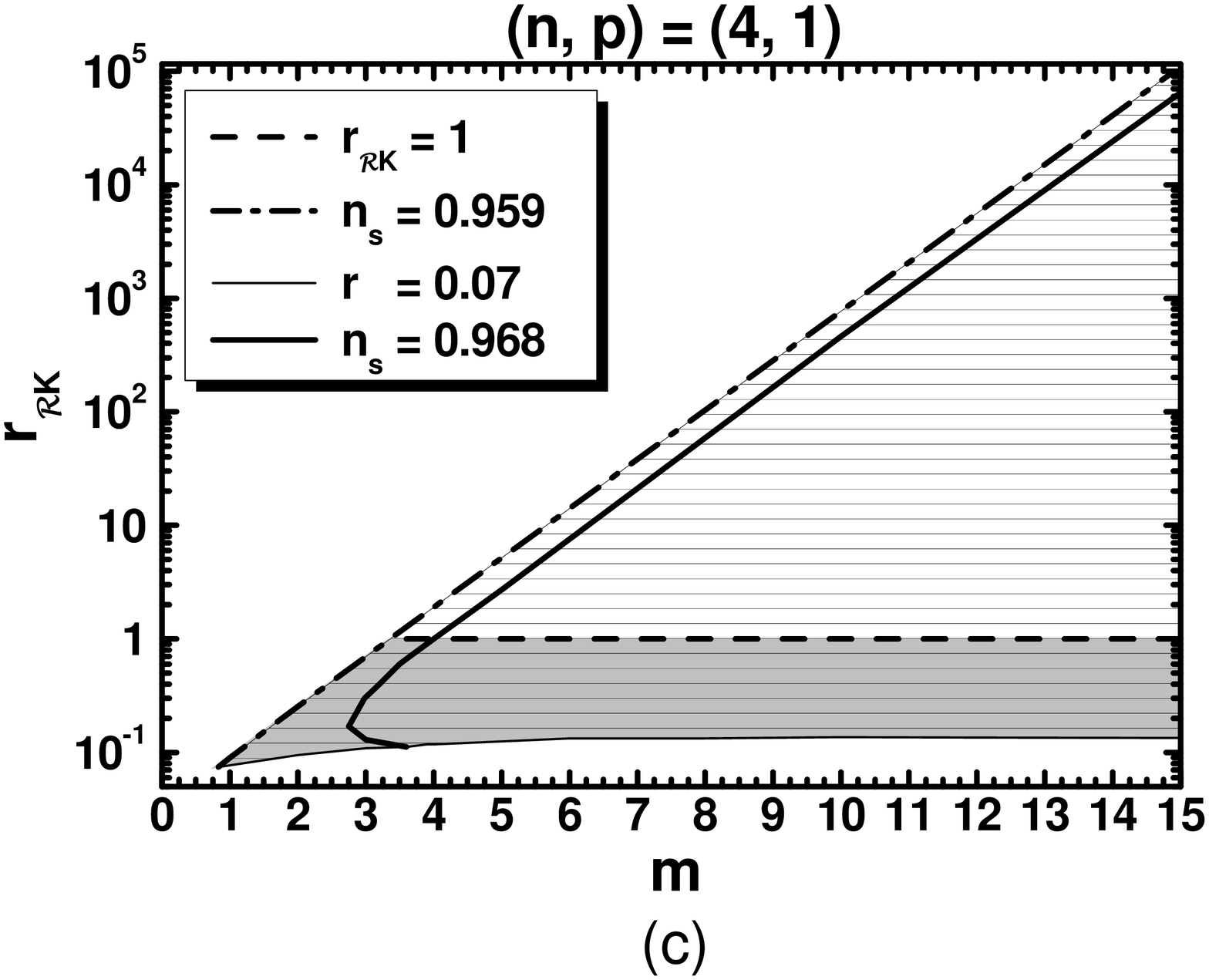,height=3.6in,angle=-90}
\hspace*{-1.2cm}
\epsfig{file=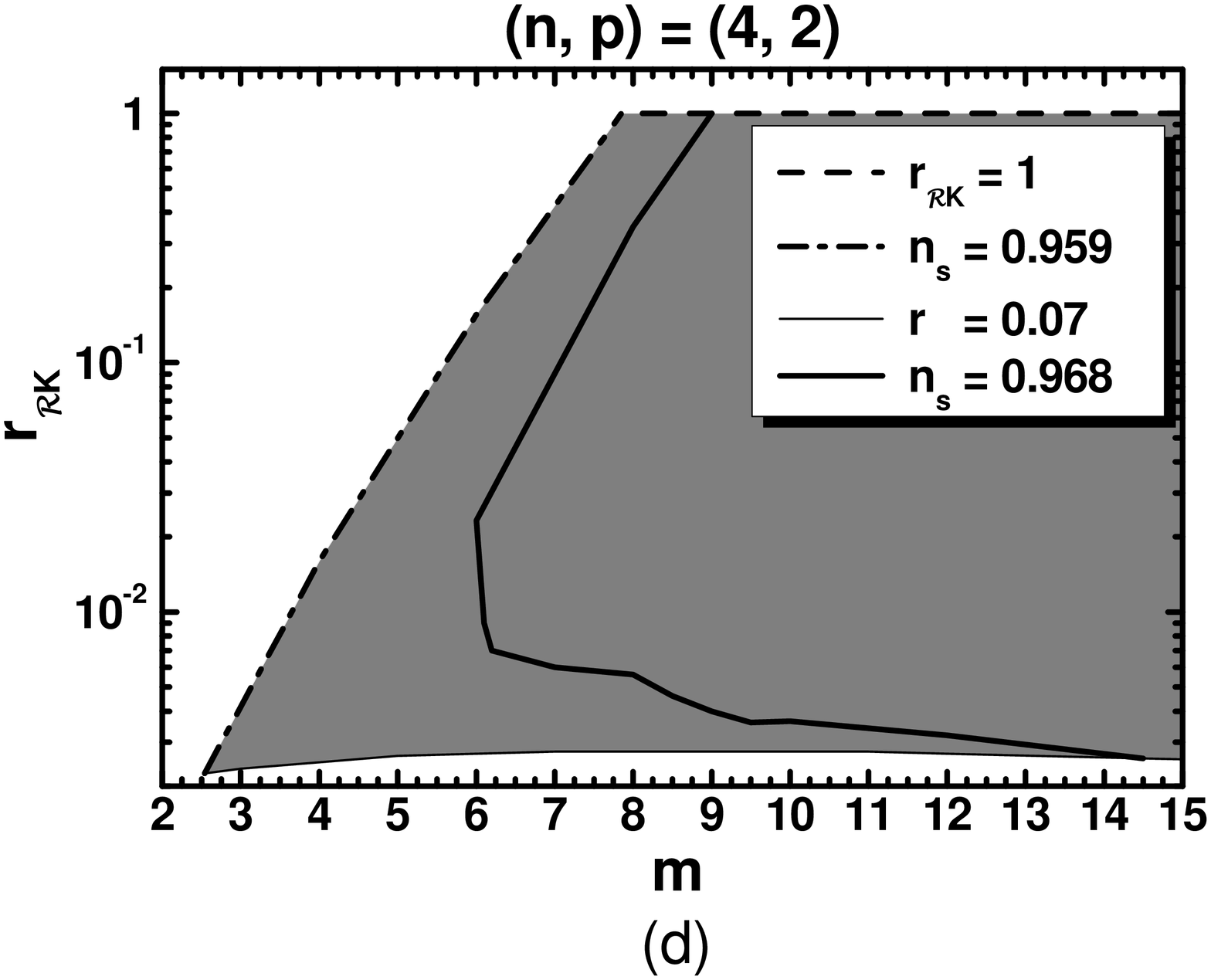,height=3.6in,angle=-90} \hfill
\end{minipage}\renewcommand{\arraystretch}{1.1}
\hfill\begin{center} {\ftn \begin{tabular}{|c||c|c|c|c|}\hline
{\sc Plot} &{{\ssz \sffamily (a)}: $(n, p)=(2,1)$}&{{\ssz
\sffamily (b)}: $(n, p) =(2,2)$}&{{\ssz \sffamily (c)}: $(n,
p)=(4, 1)$}&{{\ssz \sffamily (d)}: $(n, p) =(4, 2)$}\\\hline\hline
$m$&$4.5 - 15~\{8.2\}$&$12.7-15$&$2.75-15~\{4\}$&$6-14.5$\\
$\rs$&$0.03 - 32~\{1\}$&$0.0048-0.15$&$0.112-6.5\cdot10^4~\{1\}$&$0.0026-1$\\
$r/0.01$&$\{0.63\}~0.17-7$&$0.59-3.2$&$\{1.1\}~0.06-7$&$0.42-7$\\
$\Dex/0.1$&$\{2\}~1.1-7.5$&$1.5-4.1$&$\{1.8\}~0.4-5.6$&$1.9-5.5$\\\hline
\end{tabular}}
\end{center}\vspace*{-.08in}\renewcommand{\arraystretch}{1.}
\hfill \vchcaption{\sl\small Allowed (lightly gray and gray
shaded) regions in the $m-\rs$ plane for  {\sffamily\ssz (a)} $(n,
p)=(2, 1)$, {\sffamily\ssz (b)} $(n, p)=(2, 2)$, {\sffamily\ssz
(c)} $(n, p)=(4, 1)$ and  {\sffamily\ssz (d)} $(n, p)=(4, 2)$. For
$p=1$ the allowed regions in the panels {\sffamily\ssz (a)} and
{\sffamily\ssz (b)} are extended to the whole lined region for the
non-SUSY models and the SUGRA ones employing $K=K_i$ with
$i=1,...,7$. The conventions adopted for the various lines are
also shown. The allowed ranges of $m$, $\rs$, $r$ and $\Dex$ along
the thick solid lines in the plots are listed in the table. The
limiting values obtained imposing $\rs\leq1$ for $p=1$ are
indicated in curly brackets.}\label{fig2}
\end{figure}


Enforcing the constraints of \Sref{const} we delineate in
\sFref{fig2}{a}, {\sf\ftn (b)}, {\sf\ftn (c)} and {\sf\ftn (d)}
the allowed regions of our models for $(n,p)=(2,1), (2,2), (4,1)$
and $(4,2)$ respectively by varying continuously $\rs$ and $m$.
The conventions adopted for the various lines are also shown in
each plot. In particular, the dashed line originates from the
restriction $\rs\leq1$, and the dot-dashed and thin lines come
from the lower and upper bounds on $\ns$ and $r$, respectively --
see \Eref{nswmap}. The lined regions in \sFref{fig2}{a} and
{\ftn\sf (b)} are allowed in the non-SUSY regime and in the SUSY
one for $K=K_i$ with $i=1,...,7$ since in this case $\rs$ is
unbounded as shown in \Sref{eff}. If we use, though, the $K_i$'s
with $i=8,9$ and $10$ then only the lightly gray shaded regions
are allowed. In \sFref{fig2}{c} and {\ftn\sf (d)} where $p=2$ the
upped bound on $\rs$ is applied in both cases and so, the gray
shaded regions are allowed. Fixing $\ns$ to its central value in
\Eref{nswmap} we obtain the thick solid lines along which the
various parameters of the models range as shown in the table
displayed below the relevant plots. Let us clarify that the
maximal $r$ and $\Dex$ values corresponds to the minimal $\rs$
occurring at the intersection point of the thick and thin solid
lines whereas the minimal $r$ and $\Dex$ values corresponding to
the maximal $\rs$ are localized at the junction of the thick solid
line either with the dashed line, if the constraint $\rs\leq1$ is
applied, or with the right vertical axis at the maximal $m$ set by
hand -- see \Sref{eff}.

\begin{figure}[!t]\vspace*{-.44in}\begin{tabular}[!h]{cc}\begin{minipage}[t]{7.in}
\hspace*{.6in}
\epsfig{file=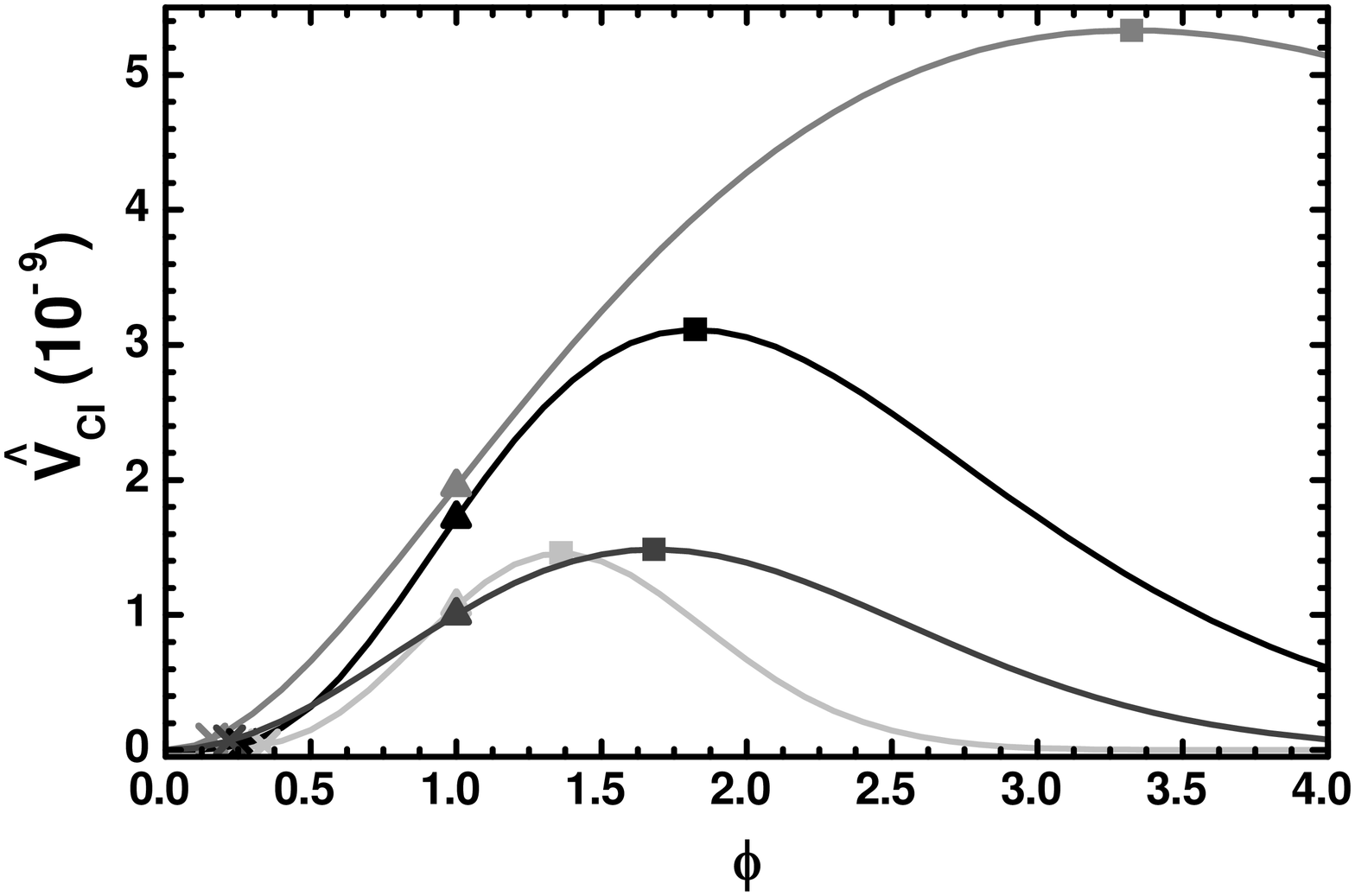,height=3.65in,angle=-90}\end{minipage}
&\begin{minipage}[h]{3.in}
\hspace{-3.5in}{\vspace*{-2.5in}\includegraphics[height=7.9cm,angle=-90]
{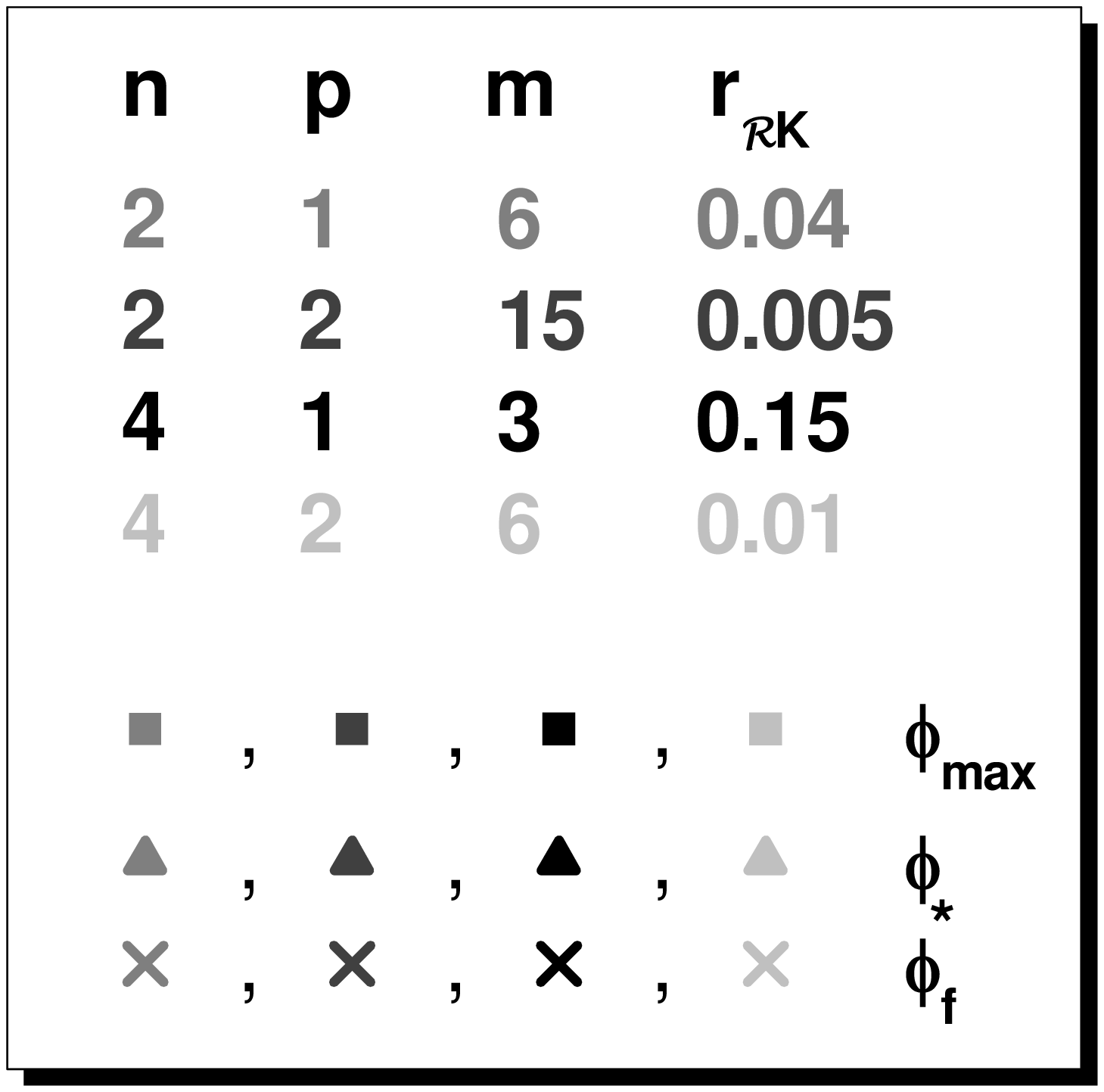}}\end{minipage}
\end{tabular}
\hfill\begin{center}\renewcommand{\arraystretch}{1.2} {\ftn
\begin{tabular}{|c|c|c|c|c||c|c|c|c|c|c|c|c|c|}\hline
\multicolumn{5}{|c||}{\sc Input
Parameters}&\multicolumn{9}{|c|}{\sc Output Parameters}\\\hline
$n$&$p$&$m$&$\rs$&$\Ns$&$\ck$&$\ld/10^{-4}$&\sex&$\sgf$&$\sg_{\rm
max}$&$\Dex$&$\ns$&$\as/10^{-4}$&$r/10^{-2}$\\\hline \hline
$2$&$1$&$6$&$0.04$&$52.4$& $56.5$&$0.084$8&$11.2$&$0.16$&$3.3$&$0.7$&$0.968$ & $-6.2$& $6.15$\\
$2$&$2$&$15$&$0.005$&$51.8$& $35.4$&$0.5$
&$9.7$&$0.22$&$1.7$&$0.4$&$0.968$
& $-6.4$& $3.1$\\
$4$&$1$&$3$&$0.15$&$57.6$& $53.5$&$1.12$&$13.4$&$0.26$&$1.8$&$0.45$&$0.968$ & $-5.6$& $5.4$\\
$4$&$2$&$6$&$0.01$&$57.4$&
$54$&$2.48$&$12.6$&$0.34$&$1.4$&$0.3$&$0.968$ & $-5.7$& $3.3$
\\\hline
\end{tabular}}
\end{center}\vspace*{-.08in} \hfill \vchcaption[]{\sl \small The inflationary
potential $\Vhi$ as a function of $\sg$ for $\sg>0$ and $(n,p,m,
\rs)=(2,1,6,0.04)$ (gray line), $(n,p,m, \rs)=(2,2,12,0.005)$
(dark gray line), $(n,p,m, \rs)=(4,1,3, 0.15)$ (black line), or
$(n,p,m, \rs)=(4,2,6, 0.01)$ (light gray line). The values of
$\sgx=1$, $\sgf$ and $\sg_{\rm max}$ are also indicated in each
case and listed together with the corresponding inflationary
observables in the table. }\label{fig3}
\end{figure}\renewcommand{\arraystretch}{1.}

From our findings in the figures above we infer that as $p$
increases the allowed regions are considerably shrunk regardless
of the range of the validity of the effective theory. Indeed, from
\Fref{fig1} we see that the curves in \sFref{fig1}{b} and
({\ftn\sf d}), where $p=2$, move to the left w.r.t their position
in \sFref{fig1}{a} and ({\ftn\sf c}), where $p=1$, and the
required $m$'s increase. This fact can be understood by
\Eref{Vmax} where we see that increasing $p$, $\sgmax$ (and
$\Dex$) decreases sharply (due to the relevant exponent) reducing,
thereby, the viable space of successful nMI. From the same formula
we infer that $\sgmax$ decreases also as $\rs$ increases for fixed
$m$ and $p$. This is the origin of the turn to the left of the
various lines in \Fref{fig1}. Similar behavior is observed in
Fig.~1-{\ftn\sf (b)} of \cref{var} for $n>0$ (in the notation of
that reference) where hilltop solutions are exhibited too.
Contrary to that set-up, though, hilltop nMI is solely attained
here. The relevant parameter $\Dex$, for central $\ns$, can be as
large as $75\%$ and remains larger than $18\%$ for $\rs\geq1$.
That is, the required tuning is not severe. Actually, this is
milder than the one obtained within the models of F-term hybrid
inflation \cite{gpp} where $\Dex\leq0.3$. The hilltop solutions
are extensively utilized there for reducing $\ns$ in the range of
\Eref{nswmap}. Moreover, our models are testable by the
forthcoming experiments \cite{ictp} searching for primordial
gravity waves since $r\gtrsim0.0059$ for $\rs\geq1$ and
$r\gtrsim6\cdot10^{-4}$ without the latter restriction. In all,
$\as$ is confined in the range $-(5-6)\cdot10^{-4}$ and so,
consistency with the fitting of data with the $\Lambda$CDM+$r$
model \cite{plin} is certainly maintained.


We complete our numerical analysis by studying the structure of
$\Vhi$ as a function of $\sg$. This is visualized in \Fref{fig3},
where we depict $\Vhi$ (gray, dark gray, black and light gray
line) versus $\sg$ for $\sgx=1$ and $(n,p,m,\rs)=(2,1,6,0.04)$,
$(2,2,15,5\cdot10^{-3})$, $(4,1,3,0.15)$ and $(4,2,6,0.01)$
respectively resulting to $\ns=0.968$. The extracted values of the
relevant parameters and observables are accommodated in the table
below the plot and compared with our semianalytical estimates in
\Sref{semi}. As anticipated in \Sref{nonsusy}, and contrary to the
picture in \crefs{nMkin, jhep,nMHkin,var} -- cf. Fig.~2 of
\crefs{jhep, var} --, $\Vhi$ possesses a clear maximum, which does
not upset, though, the realization of nMI since $\sgmax\gg\sex$
with $\Dex\geq0.3$. The relatively high $r$ values encountered
here are associated with $\sex\gg1$ in accordance with the Lyth
bound \cite{lyth}. However, this fact does not invalidate our
scenario, from the point of view of the effective theory, since
$\sgx=1$ saturating \sEref{subP}{b}. This is accomplished
selecting conveniently $\ck$ as explained in \Sref{resa}.
Actually, the $\ck$'s shown in the table of \Fref{fig3} coincide
with the lowest possible $\ck$'s since we employ the maximal
possible $\sgx$, $\sgx=1$.

Note, finally, that no attractor is pinned down in our setting
(even for $\ca\gg\ck$) and no acceptable inflationary solutions
are detected for $m=0$ in contrast to the findings of
\crefs{nMkin,jhep}. As opposed to the situation in \crefs{nMkin,
nMHkin, var} also, the $m=1$ case, which simplifies the $K_i$'s
with $i=2,4,6,7,8,9$ and $10$ is rather limited here -- see
\Fref{fig2}.

\subsection{(Semi)Analytical Results}\label{resa}

With our numerical solutions in hand, we can now derive some
(semi)analytic expressions which allow us to obtain a satisfactory
understanding of our numerics. We focus on the solutions with
$\rs\ll1$ which consist the bulk of our findings, are more natural
in the sense of the discussion in \Sref{fhi4} and ensure $r$'s of
order $0.01$ consistently with the $95\%$ c.l. region of BK14 data
-- see \Sref{nspl}. Under this basic assumption, $J$ is well
approximated by
\beq J\simeq\sqrt{{\ck}{\fr^{m-1}}}\,. \label{J1}\eeq
Obviously, $J$ is $n$ independent and for $m=1$ it becomes $\phi$
independent too leading, thereby, to exceptionally simple analytic
results -- see below. Employing \Eref{J1} the slow-roll parameters
in \Eref{srcon} can be calculated for any $m$ as follows
\beq \label{sr}\eph=\frac{\lf n -
2p\ca\sg^p\rg^2}{2\ck\sg^2\fr^{m-1}} \>\>\>\mbox{and}\>\>\>
\ith=\frac{2 n^2 - n (2 +(7 + m)p\ca\sg^p)+2p\ca\sg^p (2 +p(\ca(3
+ m)\sg^p-2))}{2\ck\sg^2\fr^{m-1}}\,\cdot \eeq
Expanding $\eph$ and $\ith$ for $\sg\ll1$ we can convince
ourselves that $\sgf\ll\sgx$. Indeed, \Eref{srcon} is saturated at
the maximal $\sg$ value, $\sgf$, from the following two values
\beqs \beq \sg_{1\rm f}\simeq\begin{cases} n\lf D_1+ \sqrt{8 +
D_1^2}\rg/4\sqrt{\ck},&\mbox{for}\>\>\>p=1\\
n/\sqrt{\ck}\sqrt{2 + D\rs},&\mbox{for}\>\>\>p=2
\end{cases}\label{sg1f}\eeq
with $D_1=(n(1-m)-4)\rs$, $D=n (8 + (m-1) n)$ and
\beq \sg_{2\rm f}\simeq \begin{cases} \lf n
D_2 + \sqrt{n}\sqrt{16 (n-1) + n D_2^2}\rg/4 \sqrt{\ck},&\mbox{for}\>\>\>p=1\\
\sqrt{n(n-1)}/\sqrt{\ck}\sqrt{1 +(4+
D)\rs},&\mbox{for}\>\>\>p=2\end{cases} \label{sg2f}\eeq\eeqs
with $D_2=(m-9+2n(1-m))\rs$. Here $\sg_{1\rm f}$ and  $\sg_{2\rm
f}$ are such that $\eph\lf\sg_{1\rm f}\rg\simeq1$ and
$\ith\lf\sg_{2\rm f}\rg\simeq1$.

Moreover, \Eref{Prob} is written as
\beq
\label{Proba}\sqrt{\As}=\frac{\ld\fr(\sgx)^{(m-3)/2}\sqrt{\ck}\sgx^{1
+ n/2}}{2^{1+{n\over4}}\sqrt{3}\pi (n-2p\ca\sgx^p)}\,\cdot \eeq
Finally, $\Ns$ can be computed from \Eref{Nhi} as follows
\begin{equation}
\label{Nhig} \Ns\simeq\int_{\sgf}^{\sgx}d\sg\:\frac{\ck\sg
e^{(m-1)\ca\sg^p}}{n - 2p\ca\sg^p}\,\cdot
\end{equation}
The implementation of the integration obliges us to single out two
cases, one for $m=1$ studied in \Sref{pm1} and one for $m\neq1$
investigated in \Sref{semi}.

\subsubsection{Analytic Results for $p=m=1$}\label{pm1}

As shown in \Fref{fig2}, there is a tiny slice of the allowed
parameter space where $p=m=1$ for both $n=2$ and $4$. In this case
$J$ becomes $\sg$ independent and $\se$ is related to $\sg$ by the
simple expression $\se=\sqrt{\ck}\sg$.  Moreover, the integration
of \Eref{Nhig} gets simplified with result
\begin{equation}
\label{Nhi1}  \Ns=\int_{\sgf}^{\sgx}\frac{d\sg\, \ck\sg}{n - 2
\ca\sg}\simeq \frac{\ck}{2\ca}\lf\frac{n}{2\ca}\ln\frac{n}{n -
2\ca\sgx}-\sgx\rg\,,\eeq
where we take into account that $\sgx\gg\sgf$. \Eref{Nhi1} can be
solved w.r.t $\sgx$ yielding
\begin{equation}
\label{sgx}\sgx\simeq
n\big(1+W_0(y_\star)\big)/{2\ca}\>\>\mbox{with}\>\>y_\star=-e^{-\lf1+4\rs^2\Ns/n\rg}.
\end{equation}
Here $W_0$ is the Lambert $W$ or product logarithmic function
\cite{wolfram}. Obviously there is a lower bound on $\ck$ for
every $\rs$ above which \sEref{subP}{b} is fulfilled. Indeed, from
\Eref{sgx} we have
\begin{equation}
\label{ckmin}\sgx\leq1~~\Rightarrow~~\ck\geq
n^2\big(1+W_0(y_\star)\big)^2/{4\rs^2}
\end{equation}
and so, our proposal can be stabilized against corrections from
higher order terms, despite the fact that $\sex\gg1$. From
\Eref{Proba} we can also derive a constraint on $\ld/\ck^{n/4}$,
i.e.,
\beq \label{lana} \ld= -2^{2 +{3 n}/{4}}n^{-{n/2}}\sqrt{3\As}\pi
\ck^{n/4} \fr(\sgx) W_0(y_\star) \big(\rs/(1 +
W_0(y_\star))\big)^{(2 +n)/2}\,. \eeq
Upon substitution of \Eref{sgx} into \Eref{ns} we find the
following expressions which assure that $\ns$ and $r$ drop as
$\rs$ increases for $n$ fixed -- cf. \Fref{fig1}. Namely
\beqs\bea\nonumber  &&\hspace*{-1cm}\ns=1 -
4\rs^2\frac{2+nW_0^2(y_\star)}{n\lf1+W_0(y_\star)\rg^2}\\
&&\hspace*{-.5cm}\simeq1 - \frac{2 + n}{2\Ns} +
\frac43\lf\frac2n\rg^{\frac12} \frac{n-1}{\Ns^\frac12}\rs -
\frac{2}{9 n}(10 + 11 n)\rs^2+ \frac{8 \sqrt{2}}{135 n^\frac32}(11
n-23)\Ns^{\frac12}\rs^3,\>\>\>\>\>\label{ns1}
\\\label{as1}
&&\hspace*{-1cm}\as=32\frac{W_0(y_\star)}{n^2}\frac{2-nW_0(y_\star)}{\lf1+W_0(y_\star)\rg^4}\rs^4\simeq-\frac{2
+ n}{2\Ns^2} + \frac{2 \sqrt{2} (n-1)}{3 (n \Ns^3)^\frac12} \rs +
\frac{4 \sqrt{2} (23 - 11 n)}{135(n^3\Ns)^\frac12} \rs^3,
\\&&\hspace*{-1cm} r=32\rs^2\frac{W_0^2(y_\star)}{\lf1+W_0(y_\star)\rg^2}\simeq\frac{4 n}{\Ns}
-\frac{32 \sqrt{2n}}{3(\Ns)^\frac12}\rs + \frac{176}9 \rs^2
-\frac{704(2\Ns)^\frac12}{135\sqrt{n}}\rs^3\,.
\label{rs1}\eea\eeqs

To appreciate the validity of our analytic estimates, we test them
against our numerical ones. We use two sets of input parameters
(for $n=2$ and $4$) and we present in Table~\ref{tab5} their
response by applying our numerical procedure (first five columns
to the right of the leftmost three ones) or using the formulae
above (next five columns). We see that the results are quite close
to each other.

\renewcommand{\arraystretch}{1.25}
\begin{table}[!t]
{\ftn \bec\begin{tabular}{|c|c|c||c|c|c|c|c||c|c|c|c|c|c|c|}\hline
\multicolumn{3}{|c||}{\sc Input}&\multicolumn{10}{|c|}{\sc Output
Parameters}\\\cline{4-13}
\multicolumn{3}{|c||}{\sc Parameters}&\multicolumn{5}{|c||}{\sc
Numerical Values}&\multicolumn{5}{|c|}{\sc Analytic
Values}\\\hline
$n$&$\rs$&$\ck$&$\ld/10^{-4}$&$\Ns$&$\ns$&$\as/10^{-4}$&$r$
&$\ld/10^{-4}$&$\Ns$&$\ns$&$\as/10^{-4}$&$r$\\\hline \hline
$2$&$0.05$&$126$&$0.97$ &$52.2$&$0.962$ & $-6.3$& $0.049$&$0.98$
&$52.5$&$0.962$ & $-6.4$&$0.049$ \\
$4$&$0.08$&$251.5$&$3.3$ &$57.5$&$0.96$ & $-6.3$& $0.067$&$3.34$
&$58$&$0.959$ & $-6.5$&$0.069$
\\\hline
\end{tabular}\eec}
\hfill \vchcaption[]{\sl\small Comparison between the numerical
and analytic results for $p=m=\sgx=1$ and two different sets of
input and output parameters of our model.} \label{tab5}
\end{table}\renewcommand{\arraystretch}{1.}

\subsubsection{Semi-analytic Results for $m\neq1$}\label{semi}

Contrary to the situation in \crefs{nMkin, nMHkin, var} the $m=1$
case is not the central one in the allowed areas -- see
\Fref{fig2}. As a consequence the purely analytic verification
above of our numerical results has to be extended to other $m$'s
too.

Using the estimation of $J$ in \Eref{J1} we can extract $\se$ as
function of $\sg$
\beq\label{se} \se=\frac1{\sqrt{(m-1)\rs}}\cdot\begin{cases} 2\big(\fr^{(m-1)/2}-1\big)/\sqrt{(m-1)\rs},& \mbox{for}~~p=1\\
\sqrt{\pi/2}\, \erfi\lf\sqrt{(m-1)\ck\rs}\sg/\sqrt{2}\rg,&
\mbox{for}~~p=2\end{cases}\eeq
where $\erfi$ is the imaginary error function \cite{wolfram}. As
we see in the table of \Fref{fig3}, the attainment of nMI requires
$\sex\gg1$, whereas \sEref{subP}{b} dictates $\sgx\leq1$.
\Eref{se} assures that both requirements above can be met since
$\se$ is increasing function of $\ck$ for fixed $\sg$. Indeed,
changing iteratively $\ck\gg1$ for fixed $\sgx\leq1$ we may obtain
any possible $\sex\gg1$.

Taking into account that $\sgx\gg\sgf$, $\Ns$ can be computed from
\Eref{Nhi} as follows
\beqs\begin{equation} \label{Nhin} \Ns= \begin{cases} -\frac1{2
\rs^2} \lf\frac{\fr(\sgx)^{m-1}}{m-1}+e^{m-1}\Ei\lf(m-1) (\ca
\sgx-1)\rg\rg+N_0^{(2,1)},
&\mbox{for}\>\>\>(n,p)=(2,1)\\-\frac{e^{(m-1)/2}}{8\rs}\Ei\lf(m-1)(2\ca\sgx^2-1)/2\rg+N_0^{(2,2)},&
\mbox{for}\>\>\>(n,p)=(2,2)\\
-\frac1{2 \rs^2}
\lf\frac{\fr(\sgx)^{m-1}}{m-1}+2e^{2(m-1)}\Ei\lf(m-1) (\ca
\sgx-2)\rg\rg+N_0^{(4,1)},& \mbox{for}\>\>\>(n,p)=(4,1)
\\-\frac{e^{m-1}}{8\rs}\Ei\lf(m-1)(\ca\sgx^2-1)\rg+N_0^{(4,2)},&
\mbox{for}\>\>\>(n,p)=(4,2)
\end{cases}
\end{equation}
where $\Ei$ is the integral exponential function \cite{wolfram}
and $N_0^{(n,p)}$ is a constant term which reads
\begin{equation} \label{N0} N_0^{(n,p)}= \frac1{8(m-1) \rs^2}\cdot\begin{cases}
4\lf1 + e^{m - 1}(m-1) \Ei(1 - m)\rg,&\mbox{for}\>\>\>(n,p)=(2,1)\\
e^{(m - 1)/2}(m-1)\rs \Ei((1 - m)/2),& \mbox{for}\>\>\>(n,p)=(2,2)\\
4\lf1 + 2e^{2(m - 1)}(m-1) \Ei(2 - 2m)\rg,& \mbox{for}\>\>\>(n,p)=(4,1)\\
e^{m - 1}(m-1) \Ei(1 - m),& \mbox{for}\>\>\>(n,p)=(4,2).
\end{cases}
\end{equation}\eeqs

\renewcommand{\arraystretch}{1.25}
\begin{table} \bec
{\ftn\begin{tabular}{|c||l|l|}\hline
$(n,p)$&$(2,1)$&$(2,2)$\\
\hline\hline
$c^{(n,p)}_{1N}$&$2m/3$& $(m+1)/2$\\
$c^{(n,p)}_{2N}$&$(m^2+1)/4$& $(m^2 + 2 m +5 )/6 $\\
$c^{(n,p)}_{3N}$&$(m^3+ 3m +2)/15$& $ (m^3+ 3 m^2 + 15 m  + 29)/24$\\
$c^{(n,p)}_{4N}$&$(m^4+ 6 m^2  + 8 m + 9)/72$& $ (m^4+ 4 m^3+ 30 m^2 + 116 m + 233)/120$\\
$c^{(n,p)}_{5N}$&$(m^5+ 10 m^3+ 20 m^2+ 45 m+44)/420$& $ (m^5+ 5 m^4 + 50 m^3+ 290 m^2+1165m+2329)/720$\\
\hline\hline
$(n,p)$&$(4,1)$&$(4,2)$\\
\hline\hline
$c^{(n,p)}_{1N}$& $(2m -1)/3$&$3c^{(2,1)}_{1N}/4$\\
$c^{(n,p)}_{2N}$& $( 2m^2-2m +1)/8$&$2c^{(2,1)}_{2N}/3$\\
$c^{(n,p)}_{3N}$& $(4m^3-6m^2+6m -1)/60$&$5c^{(2,1)}_{3N}/8$\\
$c^{(n,p)}_{4N}$& $(2m^4- 4m^3+ 6m^2-2m+1)/144$&$5c^{(2,1)}_{4N}/3$\\
$c^{(n,p)}_{5N}$& $(4m^5+10m(1-m+2m^2-m^3)+1)/1680$&$7c^{(2,1)}_{5N}/12$\\
\hline
\end{tabular}}
\end{center}
\vchcaption[]{\sl\small The coefficients $c^{(n,p)}_{jN}$ in
\Eref{Nhia} for $(n,p)=(2,1), (2,2), (4,1)$ and $(4,2)$.}
\label{tab3}
\end{table}\renewcommand{\arraystretch}{1.}

Since $\Ns$ depends on $\sgx$  in a rather complicate way, it is
not doable to solve analytically the equations above w.r.t $\sgx$
and derive formulas for the minimal possible $\ck$, and the
observables $\ns, \as$ and $r$ as functions of $\Ns$ -- cf.
Eqs.~(\ref{ckmin}), (\ref{ns1}) -- (\ref{rs1}). This difficulty
insists even if we expand $\Ns$ in (convergent) series of
$\rs\ll1$. Indeed, doing so we obtain the following expressions
\begin{equation}
\label{Nhia} \Ns=
\frac18\ck\sgx^2\sum_{j=0}^5\rs^j\cdot\begin{cases}
2c_{jN}^{(2,1)}\ck^{j/2}\sgx^j,& \mbox{for}\>\>\>(n,p)=(2,1)
\\ 2c_{jN}^{(2,2)}\ck^{j}\sgx^{2j},& \mbox{for}\>\>\>(n,p)=(2,2)
\\c_{jN}^{(4,1)}\ck^{j/2}\sgx^{j},& \mbox{for}\>\>\>(n,p)=(4,1)
\\ c_{jN}^{(4,2)}\ck^{j}\sgx^{2j},& \mbox{for}\>\>\>(n,p)=(4,2)
\end{cases}
\end{equation}
where the coefficients $c_{jN}^{(n,p)}$ are arranged in
\Tref{tab3} with $c_{0N}^{(n,p)}=1$ for any $(n,p)$. The
difficulty related to the derivation of $\sgx$ in terms of $\Ns$
can be overcome by fixing $\sgx$ to a specific value and find
(numerically) the corresponding $\ck$ in order to obtain the
required $\Ns$ by \Eref{Nhi}. This is possible since nMI can be
realized for any $\sgx\leq1$ selecting conveniently $\ck$ for
given $n, p, m$ and $\rs$. We verified that \Eref{Nhia} with fixed
$\Ns$ exhibit unique real and positive solutions $\ck=\cks$ for
$\sgx=1$. Substituting these values for $\ck$ and $\sgx$ into
\eqs{sr}{ns} we can obtain relatively simple expressions for the
inflationary observables as functions of $\cks$. Namely, we get
\beqs\bea \label{nsa}  \ns&\simeq& 1-\frac{n(2+n)}{\frs^{m-1}\cks}
- \frac{(m-5)n+4(p-1)p}{\frs^{m-1}\cks^{1-p/2}} \rs
+\frac{2(m-3)p^2}{\cks^{1-p}\frs^{m-1}}\rs^2,\\
\nonumber  \as&\simeq&\frac{2
p\cks^{p/2}\rs-n}{\cks^{2}\frs^{2(m-1)}} \Big( 2 n (2 + n)+\cks^p
\lf16 + 5 n + m ((m-6) n + 8 (p-1)) - 16 p\rg p^2\rs^2
\\\nonumber && -p\cks^{p/2}  \lf 8 + n (12 + n - m (4 + n)) - 12 p
+(m-5) n p + 4 p^2\rg\rs\\ && - 2p^3 \cks^{3 p/2} (m-3) (m-1)
\rs^3\Big),\label{asa} \\ r&\simeq&\frac{8}{\cks\frs^{m-1}}\lf
n-2p\cks^{p/2}\rs\rg^2\>\>\>\mbox{with}\>\>\>\frs=\fr(\sg=1)=e^{\ca}\,.
\label{ra}\eea\eeqs
From the expressions above we deduce that $\ns$ and $r$ decrease
as $\rs$ increases for fixed $n$ and $m$, in accordance with the
behavior of the curves in \Fref{fig1}. Finally, from \Eref{Proba}
we can find $\ld_1$ corresponding to $\cks$ from the formula
\beq \label{lda} \ld_1=2^{1 + n/4}\sqrt{3\As}\pi \frs^{(3-m)/2}\lf
n - 2p\ca\rg/\sqrt{\cks}\,.\eeq
In this result we are not able to recover the proportionality of
$\ld$ on $\ck^{n/4}$ as in \Eref{lana}, since $\sgx$ is not
expressed in terms of $\ck$ as in \Eref{sgx}.

To qualify our semianalytic treatment we compare its outputs with
the numerical ones for the inputs used in \Fref{fig3}. Our
findings are displayed in \Tref{tab4} and are in good agreement
with the values evaluated numerically and exposed in the table of
\Fref{fig3}. This is improved especially for small $\rs$'s which
ensure more accurate convergence of the expansions in \Eref{Nhia}.

\renewcommand{\arraystretch}{1.25}
\begin{table}[!t]
\bec\begin{tabular}{|c|c|c|c|c||c|c|c|c|c|c|c|}\hline
\multicolumn{5}{|c||}{\sc Input
Parameters}&\multicolumn{7}{|c|}{\sc Output Parameters}\\\hline
$n$&$p$&$m$&$\rs$&$\Ns$&$\cks$&$\ld/10^{-4}$&$\sex$&$\sgf/0.1$&$\ns$&$\as/10^{-4}$&$r/10^{-2}$\\\hline
\hline
$2$&$1$&$6$&$0.04$&$52.4$& $56.4$&$0.084$ &$11.2$&$1.5$&$0.968$ & $-6.3$& $6.2$\\
$2$&$2$&$15$&$0.005$&$51.8$& $35.8$&$0.05$ &$9.7$&$2.2$&$0.969$ & $-6$& $3$\\
$4$&$1$&$3$&$0.15$&$57.6$& $55.3$&$1.12$ &$13.3$&$2.1$&$0.971$ & $-4.8$& $4.8$\\
$4$&$2$&$6$&$0.01$&$57.$& $55.4$&$2.48$ &$12.6$&$3.2$&$0.965$ &
$-4.7$& $2.9$
\\\hline
\end{tabular}\eec
\hfill \vchcaption[]{\sl\small  Input and output parameters as
found applying our semianalytical estimates.} \label{tab4}
\end{table}\renewcommand{\arraystretch}{1.}

\section{Conclusions}\label{con}

Following the strategy in \crefs{nMkin,jhep,nMHkin,var} we
proposed a class of SUSY and non-SUSY models which support hilltop
inflation compatible with observations. The main novelty of our
present proposal is the consideration of an exponential
non-minimal coupling to gravity in \Eref{fre} apart from the
potential in \Eref{Vn} and the non-canonical kinetic mixing in
\Eref{fk} already used in the \cref{nMkin}. This setting can be
elegantly implemented in SUGRA too, employing a bilinear (for
$n=2$) or a linear-quadratic (for $n=4$) superpotential term --
see \Eref{Wn} -- and one of the \Ka s given in Eqs.~(\ref{K1}) --
(\ref{K10}). Prominent in all cases is the role of a
shift-symmetric quadratic function $\hk$ in \Eref{hr} which
remains invisible in the SUGRA scalar potential while dominates
mostly the canonical normalization of the inflaton. We specified
two functional forms for the inflaton in the $K$'s, one
logarithmic in Eqs.~(\ref{K1}) -- (\ref{K7}) and one polynomial in
Eqs.~(\ref{K8}) -- (\ref{K10}), and two stabilization mechanisms
for the non-inflaton field, one with a higher order term in
Eqs.~(\ref{K1}) -- (\ref{K4}) and (\ref{K8}), and one leading to a
$SU(2)_S/U(1)$ symmetric \Km\ in Eqs.~(\ref{K5}) -- (\ref{K7}),
(\ref{K9}) and (\ref{K10}). The only discrimination between those
models arises for $p=1$ and regards the domain where the effective
theory is valid up to $\mP$.

In all, our inflationary setting depends essentially on five free
parameters ($n, p, m$, $\ld/\ck^{n/4}$ and $\rs$) which were
constrained to natural values, imposing a number of observational
and theoretical restrictions. In particular, we investigated two
pairs of $n$ and $p$ values ($n=2, 4$ and $p=1, 2$) allowing $m$
vary from $0$ to $15$. Therefore, the extra parameter, $p$, w.r.t
those used in \cref{nMkin} assists us to enlarge the allowed
parametric space and increases the naturalness of our proposal.
Confining, e.g., $\rs$ to the range $(2.6\cdot10^{-3}-1)$, where
the upper bound does not apply to the $p=1$ case in the non-SUSY
models and the SUSY ones for $K=K_i$ with $i=1,...,7$, we
succeeded to reproduce the present data for $\ns=0.968$,
negligibly small $|\as|$ and mostly observationally reachable
$r$'s in the foreseen future -- see table of \Fref{fig2}. In all
cases, $\ld/\ck^{n/4}$ is computed enforcing \Eref{Prob} and
$\Vhi$ develops a maximum which does not disturb, though, the
implementation of hilltop nMI since the relevant tuning is mostly
rather low. Our inflationary setting can be attained with
subplanckian values of the initial (non-canonically normalized)
inflaton requiring large $\ck$'s and without causing any problem
with the perturbative unitarity. It is gratifying that our scheme
remains intact from radiative corrections and the inflationary
quantities can be estimated analytically for $p=m=1$ and
semi-analytically for the remaining cases.

Let us, finally, remark that the inflaton in our present regime
was identified with a gauge singlet field. However, our models can
be also realized very economically by a Higgs-inflaton, as in
\crefs{nmH, jhep,nMHkin,var}. I.e., the inflaton could be
represented by a Higgs field involved in the breakdown of a SUSY
grand unified theory. We expect that the results of a such
possibility will be similar to those displayed here for $n=4$. The
details of this investigation, though, could be the aim of a
forthcoming publication.

\def\prdn#1#2#3#4{{\sl Phys. Rev. D }{\bf #1}, no. #4, #3 (#2)}
\def\jcapn#1#2#3#4{{\sl J. Cosmol. Astropart.
Phys. }{\bf #1}, no. #4, #3 (#2)}

\end{document}